\documentclass[11pt]{article}
\usepackage{amssymb}
\usepackage{amsmath}
\usepackage{amsfonts}
\usepackage{graphicx}
\usepackage{float}
\usepackage{hyperref}
\usepackage{wasysym}
\usepackage{marvosym}
\usepackage[margin=1in]{geometry}
\usepackage{paralist}
\usepackage{tikz}
\usepackage{enumerate}
\usepackage{enumitem}
\usepackage{tikz-qtree}
\usepackage{caption}
\usepackage{subcaption}
\usepackage{enumerate}
\usepackage{enumitem}
\usepackage{bbm}
\usepackage{breakcites}
\usepackage{float}
\usepackage{booktabs}
\usepackage{bbm}

\usepackage{multicol}
\usepackage{authblk}
 \usepackage [autostyle, english = american]{csquotes}
 \MakeOuterQuote{"}
\usetikzlibrary{decorations.pathreplacing}
\usetikzlibrary{trees,shapes,decorations, automata, arrows}

\hypersetup{
	colorlinks=true,
	citecolor=blue,
	linkcolor=blue,
	filecolor=magenta,      
	urlcolor=cyan,
}

\allowdisplaybreaks

\usepackage[hang,flushmargin]{footmisc}
\usepackage[ruled,vlined]{algorithm2e}
\def\BibTeX{{\rm B\kern-.05em{\sc i\kern-.025em b}\kern-.08em
    T\kern-.1667em\lower.7ex\hbox{E}\kern-.125emX}}
\newcommand{\tstar}[5]{% inner radius, outer radius, tips, rot angle, options
	\pgfmathsetmacro{\starangle}{360/#3}
	\draw[#5, color = red] (#4:#1)
	\foreach \x in {1,...,#3}
	{ -- (#4+\x*\starangle-\starangle/2:#2) -- (#4+\x*\starangle:#1)
	}
	-- cycle;
}

\usetikzlibrary{decorations.pathreplacing}
\usetikzlibrary{3d,decorations.text,shapes.arrows,positioning,fit,backgrounds, positioning}
\usetikzlibrary{matrix}

\tikzset{%
	every neuron/.style={
		circle,
		draw,
		minimum size=0.5cm
	},
	neuron missing/.style={
		draw=none, 
		scale=2,
		text height=.333cm,
		execute at begin node=\color{black}$\vdots$
	},
}

\graphicspath{{img}}

\tikzset{pics/fake box/.style args={% #1=color, #2=x dimension, #3=y dimension, #4=z dimension
		#1 with dimensions #2 and #3 and #4}{
		code={
			\draw[black,ultra thin,fill=#1]  (0,0,0) coordinate(-front-bottom-left) to
			++ (0,#3,0) coordinate(-front-top-right) --++
			(#2,0,0) coordinate(-front-top-right) --++ (0,-#3,0) 
			coordinate(-front-bottom-right) -- cycle;
			\draw[black,ultra thin,fill=#1] (0,#3,0)  --++ 
			(0,0,#4) coordinate(-back-top-left) --++ (#2,0,0) 
			coordinate(-back-top-right) --++ (0,0,-#4)  -- cycle;
			\draw[black,ultra thin,fill=#1!80!black] (#2,0,0) --++ (0,0,#4) coordinate(-back-bottom-right)
			--++ (0,#3,0) --++ (0,0,-#4) -- cycle;
			\path[black,decorate,decoration={text effects along path,text={CONV}}] (#2/2,{2+(#3-2)/2},0) -- (#2/2,0,0);
		}
}}

\tikzset{pics/fake box2/.style args={% #1=color, #2=x dimension, #3=y dimension, #4=z dimension
		#1 with dimensions #2 and #3 and #4}{
		code={
			\draw[black,ultra thin,fill=#1]  (0,0,0) coordinate(-front-bottom-left) to
			++ (0,#3,0) coordinate(-front-top-right) --++
			(#2,0,0) coordinate(-front-top-right) --++ (0,-#3,0) 
			coordinate(-front-bottom-right) -- cycle;
			\draw[black,ultra thin,fill=#1] (0,#3,0)  --++ 
			(0,0,#4) coordinate(-back-top-left) --++ (#2,0,0) 
			coordinate(-back-top-right) --++ (0,0,-#4)  -- cycle;
			\draw[black,ultra thin,fill=#1!80!black] (#2,0,0) --++ (0,0,#4) coordinate(-back-bottom-right)
			--++ (0,#3,0) --++ (0,0,-#4) -- cycle;
			\path[black,decorate,decoration={text effects along path,text={LATENT}}] (#2/2,{2.75+(#3-2.75)/2},0) -- (#2/2,0,0);
		}
}}

\tikzset{pics/fake boxx/.style args={% #1=color, #2=x dimension, #3=y dimension, #4=z dimension
		#1 with dimensions #2 and #3 and #4}{
		code={
			\draw[black,ultra thin,fill=#1]  (0,0,0) coordinate(-front-bottom-left) to
			++ (0,#3,0) coordinate(-front-top-right) --++
			(#2,0,0) coordinate(-front-top-right) --++ (0,-#3,0) 
			coordinate(-front-bottom-right) -- cycle;
			\draw[black,ultra thin,fill=#1] (0,#3,0)  --++ 
			(0,0,#4) coordinate(-back-top-left) --++ (#2,0,0) 
			coordinate(-back-top-right) --++ (0,0,-#4)  -- cycle;
			\draw[black,ultra thin,fill=#1!80!black] (#2,0,0) --++ (0,0,#4) coordinate(-back-bottom-right)
			--++ (0,#3,0) --++ (0,0,-#4) -- cycle;
			\path[black,decorate,decoration={text effects along path,text={INPUT}}] (#2/2,{2+(#3-2)/2},0) -- (#2/2,0,0);
		}
}}

\tikzset{pics/fake boxo/.style args={% #1=color, #2=x dimension, #3=y dimension, #4=z dimension
		#1 with dimensions #2 and #3 and #4}{
		code={
			\draw[black,ultra thin,fill=#1]  (0,0,0) coordinate(-front-bottom-left) to
			++ (0,#3,0) coordinate(-front-top-right) --++
			(#2,0,0) coordinate(-front-top-right) --++ (0,-#3,0) 
			coordinate(-front-bottom-right) -- cycle;
			\draw[black,ultra thin,fill=#1] (0,#3,0)  --++ 
			(0,0,#4) coordinate(-back-top-left) --++ (#2,0,0) 
			coordinate(-back-top-right) --++ (0,0,-#4)  -- cycle;
			\draw[black,ultra thin,fill=#1!80!black] (#2,0,0) --++ (0,0,#4) coordinate(-back-bottom-right)
			--++ (0,#3,0) --++ (0,0,-#4) -- cycle;
			\path[black,decorate,decoration={text effects along path,text={OUTPUT}}] (#2/2,{2+(#3-2)/2},0) -- (#2/2,0,0);
		}
}}

\tikzset{pics/fake embedding/.style args={% #1=color, #2=x dimension, #3=y dimension, #4=z dimension
		#1 with dimensions #2 and #3 and #4}{
		code={
			\draw[black,ultra thin,fill=#1]  (0,0,0) coordinate(-front-bottom-left) to
			++ (0,#3,0) coordinate(-front-top-right) --++
			(#2,0,0) coordinate(-front-top-right) --++ (0,-#3,0) 
			coordinate(-front-bottom-right) -- cycle;
			\draw[black,ultra thin,fill=#1] (0,#3,0)  --++ 
			(0,0,#4) coordinate(-back-top-left) --++ (#2,0,0) 
			coordinate(-back-top-right) --++ (0,0,-#4)  -- cycle;
			\draw[black,ultra thin,fill=#1!80!black] (#2,0,0) --++ (0,0,#4) coordinate(-back-bottom-right)
			--++ (0,#3,0) --++ (0,0,-#4) -- cycle;
			\path[black,decorate,decoration={text effects along path,text={EMBEDDING}}] (#2/2,{4+(#3-4)/2},0) -- (#2/2,0,0);
		}
}}

\tikzset{pics/fake embeddingo/.style args={% #1=color, #2=x dimension, #3=y dimension, #4=z dimension
		#1 with dimensions #2 and #3 and #4}{
		code={
			\draw[black,ultra thin,fill=#1]  (0,0,0) coordinate(-front-bottom-left) to
			++ (0,#3,0) coordinate(-front-top-right) --++
			(#2,0,0) coordinate(-front-top-right) --++ (0,-#3,0) 
			coordinate(-front-bottom-right) -- cycle;
			\draw[black,ultra thin,fill=#1] (0,#3,0)  --++ 
			(0,0,#4) coordinate(-back-top-left) --++ (#2,0,0) 
			coordinate(-back-top-right) --++ (0,0,-#4)  -- cycle;
			\draw[black,ultra thin,fill=#1!80!black] (#2,0,0) --++ (0,0,#4) coordinate(-back-bottom-right)
			--++ (0,#3,0) --++ (0,0,-#4) -- cycle;
			\path[black,decorate,decoration={text effects along path,text={OUTPUT}}] (#2/2,{3+(#3-3)/2},0) -- (#2/2,0,0);
		}
}}

\tikzset{pics/fake fc/.style args={% #1=color, #2=x dimension, #3=y dimension, #4=z dimension
		#1 with dimensions #2 and #3 and #4}{
		code={
			\draw[black,ultra thin,fill=#1]  (0,0,0) coordinate(-front-bottom-left) to
			++ (0,#3,0) coordinate(-front-top-right) --++
			(#2,0,0) coordinate(-front-top-right) --++ (0,-#3,0) 
			coordinate(-front-bottom-right) -- cycle;
			\draw[black,ultra thin,fill=#1] (0,#3,0)  --++ 
			(0,0,#4) coordinate(-back-top-left) --++ (#2,0,0) 
			coordinate(-back-top-right) --++ (0,0,-#4)  -- cycle;
			\draw[black,ultra thin,fill=#1!80!black] (#2,0,0) --++ (0,0,#4) coordinate(-back-bottom-right)
			--++ (0,#3,0) --++ (0,0,-#4) -- cycle;
			\path[black,decorate,decoration={text effects along path,text={FC}}] (#2/2,{1+(#3-1)/2},0) -- (#2/2,0,0);
		}
}}

\tikzset{pics/fake unroll/.style args={% #1=color, #2=x dimension, #3=y dimension, #4=z dimension
		#1 with dimensions #2 and #3 and #4}{
		code={
			\draw[black,ultra thin,fill=#1]  (0,0,0) coordinate(-front-bottom-left) to
			++ (0,#3,0) coordinate(-front-top-right) --++
			(#2,0,0) coordinate(-front-top-right) --++ (0,-#3,0) 
			coordinate(-front-bottom-right) -- cycle;
			\draw[black,ultra thin,fill=#1] (0,#3,0)  --++ 
			(0,0,#4) coordinate(-back-top-left) --++ (#2,0,0) 
			coordinate(-back-top-right) --++ (0,0,-#4)  -- cycle;
			\draw[black,ultra thin,fill=#1!80!black] (#2,0,0) --++ (0,0,#4) coordinate(-back-bottom-right)
			--++ (0,#3,0) --++ (0,0,-#4) -- cycle;
			\path[black,decorate,decoration={text effects along path,text={UNROLL}}] (#2/2,{1+(#3-1)/2},0) -- (#2/2,0,0);
		}
}}

\tikzset{pics/fake roll/.style args={% #1=color, #2=x dimension, #3=y dimension, #4=z dimension
		#1 with dimensions #2 and #3 and #4}{
		code={
			\draw[black,ultra thin,fill=#1]  (0,0,0) coordinate(-front-bottom-left) to
			++ (0,#3,0) coordinate(-front-top-right) --++
			(#2,0,0) coordinate(-front-top-right) --++ (0,-#3,0) 
			coordinate(-front-bottom-right) -- cycle;
			\draw[black,ultra thin,fill=#1] (0,#3,0)  --++ 
			(0,0,#4) coordinate(-back-top-left) --++ (#2,0,0) 
			coordinate(-back-top-right) --++ (0,0,-#4)  -- cycle;
			\draw[black,ultra thin,fill=#1!80!black] (#2,0,0) --++ (0,0,#4) coordinate(-back-bottom-right)
			--++ (0,#3,0) --++ (0,0,-#4) -- cycle;
			\path[black,decorate,decoration={text effects along path,text={ROLL}}] (#2/2,{2+(#3-2)/2},0) -- (#2/2,0,0);
		}
}}

\tikzset{pics/fake roll/.style args={% #1=color, #2=x dimension, #3=y dimension, #4=z dimension
		#1 with dimensions #2 and #3 and #4}{
		code={
			\draw[black,ultra thin,fill=#1]  (0,0,0) coordinate(-front-bottom-left) to
			++ (0,#3,0) coordinate(-front-top-right) --++
			(#2,0,0) coordinate(-front-top-right) --++ (0,-#3,0) 
			coordinate(-front-bottom-right) -- cycle;
			\draw[black,ultra thin,fill=#1] (0,#3,0)  --++ 
			(0,0,#4) coordinate(-back-top-left) --++ (#2,0,0) 
			coordinate(-back-top-right) --++ (0,0,-#4)  -- cycle;
			\draw[black,ultra thin,fill=#1!80!black] (#2,0,0) --++ (0,0,#4) coordinate(-back-bottom-right)
			--++ (0,#3,0) --++ (0,0,-#4) -- cycle;
			\path[black,decorate,decoration={text effects along path,text={ROLL}}] (#2/2,{2+(#3-2)/2},0) -- (#2/2,0,0);
		}
}}

\tikzset{pics/fake boxhorizontal/.style args={% #1=color, #2=x dimension, #3=y dimension, #4=z dimension
		#1 with dimensions #2 and #3 and #4}{
		code={
			\draw[black,ultra thin,fill=#1]  (0,0,0) coordinate(-front-bottom-left) to
			++ (0,#3,0) coordinate(-front-top-right) --++
			(#2,0,0) coordinate(-front-top-right) --++ (0,-#3,0) 
			coordinate(-front-bottom-right) -- cycle;
			\draw[black,ultra thin,fill=#1] (0,#3,0)  --++ 
			(0,0,#4) coordinate(-back-top-left) --++ (#2,0,0) 
			coordinate(-back-top-right) --++ (0,0,-#4)  -- cycle;
			\draw[black,ultra thin,fill=#1!80!black] (#2,0,0) --++ (0,0,#4) coordinate(-back-bottom-right)
			--++ (0,#3,0) --++ (0,0,-#4) -- cycle;
			\path[black,decorate,decoration={text effects along path,text={EMBEDDING}}] (0.25,#3/2,0) -- ({#2},#3/2,0);
		}
}}

\tikzset{pics/fake empty/.style args={% #1=color, #2=x dimension, #3=y dimension, #4=z dimension
		#1 with dimensions #2 and #3 and #4}{
		code={
			\draw[black,ultra thin,fill=#1]  (0,0,0) coordinate(-front-bottom-left) to
			++ (0,#3,0) coordinate(-front-top-right) --++
			(#2,0,0) coordinate(-front-top-right) --++ (0,-#3,0) 
			coordinate(-front-bottom-right) -- cycle;
			\draw[black,ultra thin,fill=#1] (0,#3,0)  --++ 
			(0,0,#4) coordinate(-back-top-left) --++ (#2,0,0) 
			coordinate(-back-top-right) --++ (0,0,-#4)  -- cycle;
			\draw[black,ultra thin,fill=#1!80!black] (#2,0,0) --++ (0,0,#4) coordinate(-back-bottom-right)
			--++ (0,#3,0) --++ (0,0,-#4) -- cycle;
		}
}}

\tikzset{circle dotted/.style={dash pattern=on .05mm off 2mm,
		line cap=round}}

\usepackage{pgfplots}
\pgfplotsset{compat=1.16}

\tikzstyle{data} = [rectangle, rounded corners, minimum width=4.5cm, minimum height=1cm,text centered, text width = 4.5cm, draw=black, fill=orange!30]
\tikzstyle{io} = [trapezium, trapezium left angle=70, trapezium right angle=110, minimum width=3cm, minimum height=1cm, text centered, draw=black, fill=violet!30]
\tikzstyle{process} = [rectangle, minimum width=4.5cm, minimum height=1cm, text centered, draw=black, text width = 4.5cm, fill=violet!30]
\tikzstyle{decision} = [diamond, minimum width=4.5cm, minimum height=3cm, text centered, text width=4.5cm, draw=black, fill=green!30]
\tikzstyle{decisiont} = [diamond, minimum width=3cm, minimum height=3cm, text centered, text width=3cm, draw=black, fill=orange!30]
\tikzstyle{decisiong} = [diamond, minimum width=3cm, minimum height=3cm, text centered, text width=3cm, draw=black, fill=green!30]
\tikzstyle{arrow} = [thick,->,>=stealth]
\tikzstyle{datae} = [rectangle, rounded corners, minimum width=4.5cm, minimum height=1cm,text centered, text width = 4.5cm, draw=black, fill=green!30]
\tikzstyle{processt} = [rectangle, minimum width=4.5cm, minimum height=1cm, text centered, draw=black, text width = 4.5cm, fill=orange!30]
\tikzstyle{processg} = [rectangle, minimum width=4.5cm, minimum height=1cm, text centered, draw=black, text width = 4.5cm, fill=green!30]

\makeatletter
\DeclareRobustCommand{\rvdots}{%
	\vbox{
		\baselineskip4\p@\lineskiplimit\z@
		\kern-\p@
		\hbox{.}\hbox{.}\hbox{.}
}}
\makeatother

\usepackage{import}
\usetikzlibrary{quotes,arrows.meta}
\usetikzlibrary{positioning}

\def\edgecolor{rgb:blue,4;red,1;green,4;black,3}

\usepackage{Ball}
\usepackage{Box}
\usepackage{RightBandedBox}

\usetikzlibrary{positioning}
\usetikzlibrary{3d} %for including external image 

\usepackage{theorem}
\newtheorem{attribute}{Attribute}
\usepackage{enumitem}
\setlist{nosep}
\usepackage{tabularx}
\newlist{steps}{enumerate}{1}
\setlist[steps, 1]{label = step \arabic*, leftmargin=5.1em}
\usepackage[colorinlistoftodos]{todonotes}

\usepackage{soul}
\usepackage{soulutf8}
\usepackage{makecell}

\begin{document}
	
    % !TeX spellcheck = en_US  

\title{Geographic ratemaking with spatial embeddings}
\author[1,3]{Christopher Blier-Wong}
\author[1,3,4]{Hélène Cossette}
\author[2,3]{Luc Lamontagne}
\author[1,3,4]{Etienne Marceau\thanks{Corresponding author: Etienne Marceau, etienne.marceau@act.ulaval.ca}}
\affil[1]{École d'actuariat, Université Laval, Québec, Canada}
\affil[2]{Département d'informatique et de génie logiciel, Université Laval, Québec, Canada}
\affil[3]{Centre de recherche en données massives, Université Laval, Québec, Canada}
\affil[4]{Centre interdisciplinaire en modelisation mathématique, Université Laval, Québec, Canada}
\date{\today}
\setcounter{Maxaffil}{0}
\renewcommand\Affilfont{\itshape\small}

\maketitle

\begin{abstract}
	Spatial data is a rich source of information for actuarial applications: knowledge of a risk's location could improve an insurance company's ratemaking, reserving or risk management processes.	Insurance companies with high exposures in a territory typically have a competitive advantage since they may use historical losses in a region to model spatial risk non-parametrically. Relying on geographic losses is problematic for areas where past loss data is unavailable. This paper presents a method based on data (instead of smoothing historical insurance claim losses) to construct a geographic ratemaking model. In particular, we construct spatial features within a complex representation model, then use the features as inputs to a simpler predictive model (like a generalized linear model). Our approach generates predictions with smaller bias and smaller variance than other spatial interpolation models such as bivariate splines in most situations. This method also enables us to generate rates in territories with no historical experience.
\end{abstract}

\section{Introduction}

Insurance plays a vital role in protecting customers from rare but costly events. Insurance companies accept to cover a policyholder's peril in exchange for a fixed premium. For insurance costs to be fair, customers must pay premiums corresponding to their expected future costs. Actuaries accomplish this task by segmenting individuals in similar risk profiles and using historical data from these classes to estimate future costs. Advances in computation and statistical learning, along with a higher quantity of available information, drive insurance companies to create more individualized risk profiles.

An important factor that influences insurance risk is where a customer lives. Locations impact socio-demographic perils like theft (home and auto insurance), irresponsible driving (auto insurance) or quality of home maintenance (e.g., if homeowners replace the roofing regularly). Natural phenomena such as weather-based perils (flooding, hail, and storms) depend on natural factors such as elevation and historic rainfall. Geographic ratemaking attempts to capture geographic effects within the rating model. Historically, actuaries use spatial models to perform geographic ratemaking.

One may think that one must include a geographic component for a model to capture geographic effects: either depending on coordinates or on indicator variables that identify a territory. Indeed, the related research from actuarial science uses the latter approach. These models require a large quantity of data to learn granular geographic effects and do not scale well to portfolios of large territories. Until we model the geographic variables that generate the geographic risks, it is unfeasible to model postal code level risk in a country-wide geographic model. In the present paper, we propose a method to construct geographic features that capture the relevant geographic information to model geographic risk. We find that a model using these features as input to a GLM can model geographic risk more accurately and more parsimoniously than previous geographic ratemaking models. 

In this paper, we construct geographic features from census data. The intuition through this paper is that since \emph{people} generate \emph{risk}, the geographic distribution of the \emph{population} (as captured by census data) relates to the geographic distribution of \emph{risk}. For this reason, we place our emphasis on constructing a model that captures the geographic distribution of the \emph{population}, and use the results from the \emph{population} model to predict the geographic distribution of \emph{risk}. If we capture the geographic characteristics of the \emph{population} correctly, then a ratemaking model using the geographic distribution of the \emph{population} as input may not require any geographic component (coordinate or territory) since the geographic distribution of the \emph{population} will implicitly capture some of the geographic distribution of \emph{risk}. We focus on the geographic distribution of \emph{populations} as an intermediate step of the predictive model. The main reason for this is that information about \emph{populations} is often free, publicly available and smooth, while information about \emph{risk} is expensive, private and noisy. 

\subsection{Spatial models}

Spatial statistics is the field of science that studies spatial models. A typical problem in spatial statistics is to sample continuous variables at discrete locations in a territory and predict the value of this variable at another location within the same territory, called spatial interpolation. The prevalent theory of spatial interpolation is \textit{regionalized variable theory}, which assumes that we can deconstruct a spatial random variable into a local structured mean, a local structured covariation, and global unstructured variance (global noise) \cite{matheron1965variables, wackernagel2013multivariate}. To compute the pure premium of an insurance contract, it suffices to capture the local mean. Simple methods like local averaging or bivariate smoothing (local polynomial regression or splines) can compute this local mean.

A very common spatial interpolation model is called kriging \cite{cressie1990origins}, which performs spatial interpolation with a Gaussian process parametrized by prior covariances. These covariances depend on variogram models, a tool to measure spatial autocorrelation between pairs of points as a function of distance. In our experience, variograms can be difficult to estimate in actuarial science on claims data due to the large volume of zero observed losses. %Related to interpolation is spatial prediction, the task of predicting the value of a random variable in a territory for which we have no previous observations \cite{miller2004tobler}. 

For risk management purposes, it is beneficial to study how geographic variables interact with each other. For instance, one could study the distribution of losses for an insurance contract conditional on the fact that nearby policyholder incurred a loss. Spatial autocorrelation models study the effect of \textit{nearness} on geographic random variables \cite{getis2010spatial}. Spatial autoregressive models, which capture the spatial effects with latent spatial variables, are common approaches; see \cite{cressie2015statistics} for details.

\subsection{Literature review}\label{ss:review}

We now review the literature of geographic ratemaking in actuarial science. One can deconstruct the spatial modeling process in three steps: 
\begin{steps}[leftmargin=5.1em]
	\item\label{item:spatial1} data preparation and feature engineering;
	\item\label{item:spatial2} main regression model;
	\item\label{item:spatial3} smoothing model or residual correction.
\end{steps}
Early geographic models in actuarial science were correction models that smoothed the residuals of a regression model, i.e., capturing geographic effects \textit{after} the main regression model, in a smoothing model (\ref{item:spatial3}). Notable examples include \cite{taylor1989use}, \cite{boskov1994premium} and \cite{taylor2001geographic}. If we address the geographic effects \textit{during} or \textit{before} the main regression model, then the smoothing \ref{item:spatial3} is not required. \cite{dimakos2002bayesian} propose a Bayesian model that captures geographic trend and dependence simultaneously to the main regression model, during \ref{item:spatial2}. This model was later refined and studied as conditional autoregressive models by \cite{gschlossl2007spatial} and \cite{shi2017territorial}. Another approach is spatial interpolation, that capture geographically varying intercepts of the model. Examples include \cite{fahrmeir2003generalized, denuit2004non, wang2017geographical, henckaerts2018data}, and other spatial interpolation methods like regression-kriging \cite{hengl2007regression}. These methods use the geographic coordinates of the risk along with multivariate regression functions to capture geographic trend. 

The above models capture geographic effects directly and non-parametrically, increasing model complexity and making estimation difficult (increasing the number of parameters, making them less competitive when comparing models based on criteria that penalize model complexity). As a result, geographic smoothing methods adjusted on residuals \ref{item:spatial3} are still the prevalent geographic methods in practice for geographic ratemaking.

In the present paper, we take a fundamentally different approach, capturing the geographic effects during the feature engineering of \ref{item:spatial1}. Instead of capturing geographic effects non-parametrically with geographic models, we introduce geographic data in the ratemaking model. Geographic data is "data with implicit or explicit reference to a location relative to the Earth" \cite{ISO:19109}. Geographic data can describe natural variables describing the
	ecosystem % Systeme entre la vie et l'environement, Biomes
	and the landform of a location, % TOPOGRAPHIE: terrain, body of water
	or artificial variables describing human settlement and infrastructure. We study the effectiveness of automatically extracting useful representations of geographic information with representation learning, see \cite{bengio2013representation} for a review of this field of computer science. 

Early geographic representation models started with a specific application, then constructed representations useful for their applications. These include \cite{eisenstein2010latent, cocos2017language} for topical variation in text, \cite{yao2017sensing} to predict land use, \cite{xu2020venue2vec} for user location prediction, and \cite{jeawak2019embedding, yin2019gps2vec} for geo-aware prediction. More recent approaches aim to create general geographic embeddings. These include \cite{saeidi2015lower}, who use principal component analysis and autoencoders to compress census information. In \cite{fu2019efficient}, \cite{zhang2019unifying} and \cite{du2019beyond}, the authors use point-of-interest and human mobility data to construct spatial representations of spatial regions in a framework that preserves intra-region geographic structures and spatial autocorrelation. The authors use graph convolutional neural networks to train the representations. In \cite{jenkins2019unsupervised}, the authors propose a method to create representations of regions using satellite images, point-of-interest, human mobility and spatial graph data. Then, \cite{blier2020encoding} propose a method that captures the geographic nature of data into a geographic data cuboid, using convolutional neural networks to construct representations using Canadian census information. The proposed model is coordinate-based, therefore more flexible than regions. \cite{hui2020predicting} use graphical convolutional neural networks to compress census data in the United States to predict economic growth trends. Finally, \cite{wang2020urban2vec} propose a framework to learn neighborhood embeddings from street view images and point-of-interest data. 

%\subsection{Overview of this paper}
%
%The remainder of this paper is structured as follows. In Section \ref{sec:spatial-rep}, we explain how we can capture spatial models' desirable properties within representations models. Section \ref{sec:crae} presents a spatial representation model, while Section \ref{sec:implementation} provides the details of an implementation on Canadian census data. We present applications of the spatial embeddings on insurance-related datasets in Section \ref{sec:application}, comparing spatial embeddings to an existing spatial model. Section \ref{sec:conclusion} concludes the paper. 

    % !TeX spellcheck = en_US
\section{Spatial representations}\label{sec:spatial-rep}

In \cite{blierwong2021rethinking}, we propose a framework for actuarial modeling with emerging sources of data. This approach separates the modeling process into two steps: a representation model and a predictive model. Using this deconstruction, we can leverage modern machine learning models' complexity to create useful representations of data and use the resulting representations within a simpler predictive model (like a generalized linear model, GLM). This section presents an overview of representation models, with a focus on spatial embeddings. 

\subsection{Overview of representation models}\label{sec:representations}

\begin{figure}[ht]
	\centering
	%\resizebox{0.5\textwidth}{!}{
%	\begin{tikzpicture}
%	\tikzstyle{connection}=[ultra thick,every node/.style={sloped,allow upside down},draw=\edgecolor,opacity=0.7]
%	\tikzstyle{copyconnection}=[ultra thick,every node/.style={sloped,allow upside down},draw={rgb:blue,4;red,1;green,1;black,3},opacity=0.7]
%	
%%	\path[use as bounding box] (0, -2.5) rectangle (20, 4);
%	
%	
%	\pic[shift={(0,0,0)}] at (0,0,0) 
%	{Box={
%			name=encoder,
%			caption=Encoder,
%			xlabel={{, }},
%			zlabel=\Large,
%			fill=red,
%			opacity=0.25,
%			height=16,
%			width=8,
%			depth=16
%		}
%	};
%	
%	
%	
%	\pic[shift={(2.5,0,0)}] at (encoder-east) 
%	{Box={
%			name=embedding,
%			caption=Embedding,
%			xlabel={{,}},
%			zlabel=,
%			fill=violet,
%			opacity=0.5,
%			height=1,
%			width=1,
%			depth=8
%		}
%	};
%	
%	\pic[shift={(2.5,0,0)}] at (embedding-east) 
%	{Box={
%			name=decoder,
%			caption=Decoder,
%			xlabel={{, }},
%			zlabel=,
%			fill=blue,
%			opacity=0.25,
%			height=16,
%			width=8,
%			depth=16
%		}
%	};
%	
%	\end{tikzpicture}}

\resizebox{\textwidth}{!}{
	\begin{tikzpicture}[node distance=6cm]
	\node (data) [data] {Input};	
	\node (encoder) [rectangle, minimum width=4.5cm, minimum height=1cm,text centered, text width = 4.5cm, draw=black, fill=red!25, right of = data] {Encoder};
	\node (embedding) [rectangle, rounded corners, minimum width=4.5cm, minimum height=1cm,text centered, text width = 4.5cm, draw=black, fill=violet!30, right of = encoder] {Embedding};
	\node (decoder) [rectangle, minimum width=4.5cm, minimum height=1cm,text centered, text width = 4.5cm, draw=black, fill=blue!25, right of = embedding] {Decoder};
	\node (output) [rectangle, rounded corners, minimum width=4.5cm, minimum height=1cm,text centered, text width = 4.5cm, draw=black, fill=orange!30, right of = decoder] {Output};
	
	\node (predict) [processt, right of = embedding, yshift= -1.5cm] {Predictive model};
	
	\draw [arrow] (data) -- (encoder);
	\draw [arrow] (encoder) -- (embedding);
	\draw [arrow] (embedding) -- (decoder);
	\draw [arrow] (decoder) -- (output);
	\draw [arrow] (embedding) |- (predict);
	
	\end{tikzpicture}
}
	\caption{Encoder/decoder architecture}\label{fig:enc-dec}
\end{figure}
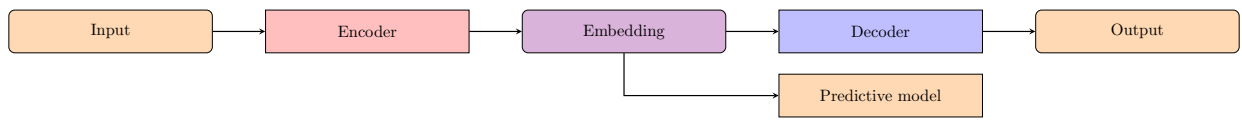

The defining characteristic of a representation model (as opposed to an end-to-end model) is that representation models never use the response variable of interest during training, relying instead on the input variables themselves or other response variables related to the insurance domain. We propose using encoder/decoder models to construct embeddings, typically with an intermediate representation (embedding) with a smaller dimension than the input data; see Figure \ref{fig:enc-dec} for the outline of a typical process. When training the representation model, we adjust the parameters of the encoder and the decoder, two machine learning algorithms. To construct the simpler regression model (like a GLM), one extracts embedding vectors for every observation and stores them in a design matrix.  

The representation construction process has four steps:
\begin{steps}
	\item\label{item:representation1} Construct an encoder, transforming input variables into a latent embedding vector.
	\item\label{item:representation2} Construct a decoder, which will determine which information the representation model must capture.
	\item\label{item:representation3} (Optional) Combine features from different sources into one combined embedding.
	\item\label{item:representation4} (Optional) Evaluate different embeddings, validate the quality of representations.
\end{steps} 
%The general advantages of our approach include transforming categorical variables into dense and compact vectors, reducing data dimension, generating reusable representations, learning non-linear transformations and interactions, and decreasing regression models' complexity.\todo[color=green!40]{L: Élaborer, une phrase pour chaque avantage}
We now enumerate a few general advantages of the representation approach. First, representation models can transform categorical variables into dense and compact vectors, which is particularly important for territories since the cardinality of this category of variables is typically very large. Second, representation models can reduce the input variable dimension into one of our choosing: we typically select an embedding dimension much smaller than the original. Third, we can build representations such that they are useful for general tasks, so we can reuse representations. Fourth, representations can learn non-linear transformations and interactions automatically, eliminating the need to construct features by hand. Finally, when using a representation model, one can reduce the regression model's complexity. If the representation model learns all useful interactions and non-linear transformations, a simple model like a GLM could replace more complex models like end-to-end gradient boosting machines or neural networks.

\subsection{Motivation for geographic embeddings}\label{ss:motivation}

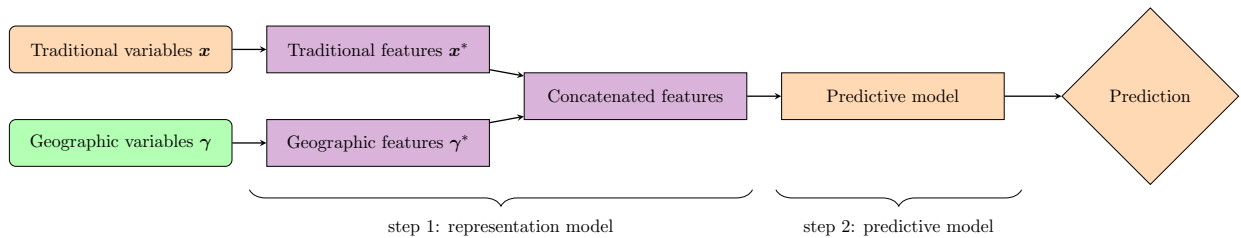
\begin{figure}[ht]
	\centering
	\resizebox{\textwidth}{!}{
	\begin{tikzpicture}[node distance=2cm]
		\node (data) [data] {Traditional variables $\boldsymbol{x}$};
		\node (emerging1) [datae,below of=data] {Geographic variables $\boldsymbol{\gamma}$};
		
		\node (representation0) [process, right of = data, xshift=3.5cm] {Traditional features $\boldsymbol{x}^*$};
		\node (representation1) [process,right of = emerging1, xshift=3.5cm] {Geographic features $\boldsymbol{\gamma}^*$};
		
		\node (representationc) [process,right of = representation1, xshift = 3.5cm, yshift=1cm] {Concatenated  features};
		
		\node (predict) [processt, right of = representationc, xshift= 3.5cm] {Predictive model};
		\node (response) [decisiont, right of = predict, xshift = 3.5cm] {Prediction};
		
		\draw [arrow] (data) -- (representation0);
		\draw [arrow] (emerging1) -- (representation1);
		\draw [arrow] (representation0) -- (representationc);
		\draw [arrow] (representation1) -- (representationc);
		\draw [arrow] (representationc) -- (predict);
		\draw [arrow] (predict) -- (response);
		
		\draw [decorate,decoration={brace,amplitude=10pt,mirror, raise = 0ex}] (2.75,-3) -- (13.5, -3) node[midway,yshift=-0.8cm]{step 1: representation model};
		\draw [decorate,decoration={brace,amplitude=10pt,mirror, raise = 0ex}] (14,-3) -- (19.25, -3) node[midway,yshift=-0.8cm]{step 2: predictive model};
		
%		\draw [decorate,decoration={brace,amplitude=10pt,mirror, raise = 0ex}] (2.75,-4) -- (19.25, -4) node[midway,yshift=-0.8cm]{End-to-end model};
		
	\end{tikzpicture}
}
	\caption{Proposed geographic ratemaking process}\label{fig:process}
\end{figure}

The geographic methods proposed in actuarial science (see the literature review) address the data's geographic nature at \ref{item:spatial2} and \ref{item:spatial3} of geographic modeling. Geographic embeddings are a fundamentally different approach to geographic models studied in actuarial science. We first transform geographic data into geographic embedding vectors, during feature engineering (\ref{item:representation1}). By using geographic embeddings in the main regression model, we capture the geographic effects and the traditional variables at the same time. Figure \ref{fig:process} provides an overview of the method. The representation model takes geographic data as input to automatically create geographic features (geographic embeddings). Sections \ref{sec:crae} and \ref{sec:implementation} respectively present architectures and implementations of geographic embedding models. Then, we combine the geographic embeddings with other sources of data (for example, traditional actuarial variables like customer age). Finally, we use the combined feature vector within a predictive model. 

The representation model's complexity does not affect the predictive model's since the representation model is \textit{unsupervised} with respect to the predictive task. We can construct representation models that are highly non-linear with architectures that capture the unique characteristics of different data sources. This model will induce geographic effects into embeddings while capturing the desirable characteristics of geographic models. In most cases, regression models (GLMs) using these geographic embeddings as inputs will have lower variance and often lower bias than geographic models, as demonstrated with examples in Section \ref{sec:application}.

In \cite{blierwong2021rethinking}, the authors present a collection of tools to construct useful actuarial representations. Section 7 of that paper deals with geographic representation ideas. The present paper aims to pursue our investigation on geographic representations for actuarial science by providing an implementation and an application. 

The representation learning framework enables one to select an architecture that captures specific desirable attributes from various data sources. There is one generally accepted desirable attribute for geographic models, called Tobler's first law (TFL) of geography. We also propose two new attributes that make geographic embeddings more useful, based on our experience with geographic models. Below is a list of these three desirable attributes for geographic embeddings.

\begin{attribute}[TFL]\label{att:tfl1}
	Geographic embeddings must follow Tobler's first law of geography. \emph{As mentioned in \cite{tobler1970computer}, "everything is related to everything else, but near things are more related than distant things". This attribute is at the core of spatial autocorrelation statistics. Spatial autocorrelation is often treated as a confounding variable, but these variables constitute information since it captures how territories are related (see, e.g.,  \cite{miller2004tobler} for a discussion). A representation model, capturing the latent structure of underlying data, generates geographic embeddings.}
\end{attribute}

\begin{attribute}[coordinate]\label{att:coordinate2}
	Geographic embeddings are coordinate-based. \emph{A coordinate-based model depends only on the risk's actual coordinates and its relation to other coordinates nearby. Polygon-based models highly depend on the definition of polygon boundaries, which could be designed for tasks unrelated to insurance.}
	
	\emph{We motivate this attribute with an example. Consider four customers A, B, C and D, that have home insurance for their house in the island of Montréal, Québec, Canada. Figure \ref{fig:dist-points-2} identifies each coordinate on a map. We also included the borders of polygons, represented by bold black lines. These polygons correspond to Forward Sortation Areas (FSA), a unit of geographic coordinate in Canada (further explained in Section \ref{sec:implementation}). Observations A and B belong to the same polygon, while observations B and C belong to different ones. However, B is closer to C than to A. Polygon-based representation models and polygon-based ratemaking models assume that the same geographic effect applies to locations A and B, while different geographic effects apply to locations B and C. However, following TFL, one expects the geographic risk between customers B and C to be more similar than the geographic risk between A and B. }	% Therefore, the issue is that the difference in geographic risk between B and C is the same as between A and D.
	
	\emph{There are two other issues with the use of polygons. The first is that the actual shapes of polygons could be designed for purposes that are not relevant for capturing geographic risk (for example, postal codes are typically optimized for mail distribution). If an insurance company uses polygons for geographic ratemaking, it is crucial to verify that risks within each polygons are geographically homogeneous. The second issue is that polygon boundaries may change in time. If postal codes are split, moved, merged or created, the geographic ratemaking model will also split, move, merge the territories, and be unable to predict for newly created postal codes. A customer living at the same address could see its insurance premium drastically increase or decrease due to the arbitrary change in polygon boundary, even if there is no actual change in the geographic risk.}
	
	\emph{Finally, the type of location information (coordinate or polygon) for the ultimate regression task may be unknown while training the embeddings. For this reason, one should not make the polygon depend on a specific set of boundary polygon shapes. On the other hand, one can easily aggregate coordinates into the desired location type for the ultimate regression task.}
	
	\begin{figure}[ht]
		\centering
		\includegraphics[width=0.7\linewidth]{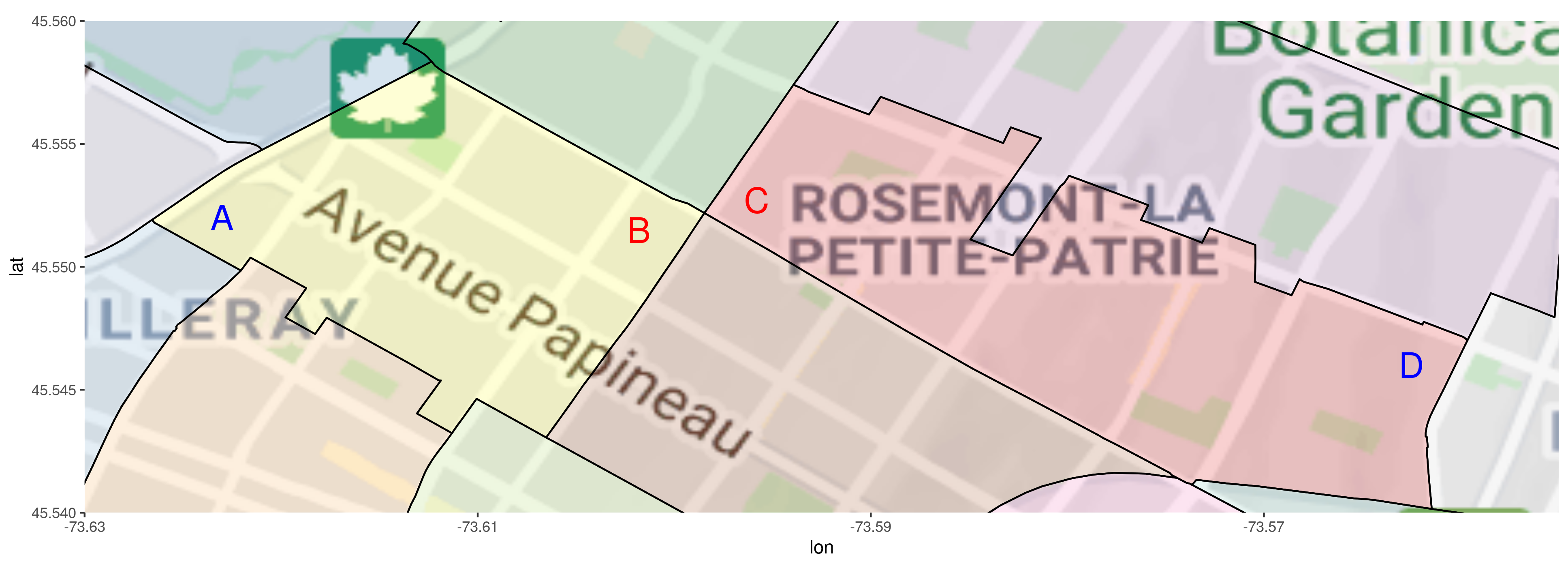}
		\caption{Coordinates vs polygons}
		\label{fig:dist-points-2}
	\end{figure}
	
\end{attribute}

\begin{attribute}[external]\label{att:external3}
	Geographic embeddings encode relevant external information. \emph{There are two reasons for using external information. The first is that geographic effects are complex, so we need a large volume of geographic information to capture the confounding variable that generates them. Constructing geographic embeddings with a dataset external to one's specific problem may increase the quantity and quality of data, providing new information for spatial prediction. The second reason is that geographic embeddings can produce rates in territories with no historical loss experience, as long as we have external information for the new territory. Traditional geographic models capture geographic effects from experience but are useless to rate new territories. If we train geographic embeddings on external datasets that cover a wider area than the loss experience of an insurance company, we may use the geographic embeddings to capture geographic effects in territories with no loss experience. This second reason is related to using one-hot encodings of territories; see \cite{blierwong2021rethinking} for further illustrations. Finally, the external information should be relevant to the insurance domain. Although geographic embeddings could be domain agnostic (general geographic embeddings for arbitrary tasks), our ultimate goal is geographic ratemaking, so we select the external information that is related to causes of geographic risks.}%\todo[color=cyan!40]{Mentionner le blogues, et quels constributions satisfaient quels attributs.}
\end{attribute}

To construct the geographic representations, we use a flexible family of machine learning methods called deep neural networks. In the past few years, actuarial applications of neural networks are increasing in popularity; see \cite{richman2020aia, blier2021machine} for recent reviews. We construct the neural networks such that the representations satisfy the three desirable attributes of geographic embeddings, so any predictive model (even linear ones) using the geographic embeddings as input will also satisfy the desirable attributes. 

\subsection{Relationship with word embeddings}

The geographic embeddings approach that we propose is largely inspired by word embeddings in natural language processing (NLP). The fundamental idea of word embeddings dates back to linguists in the 1950's. On the subject of the distributional hypothesis and synonyms, \cite{harris1954distributional} states "If A and B have some environments in common and some not $\dots$ we say that they have different meanings, the amount of meaning difference corresponding roughly to the amount of difference in their environments." Although this \emph{environment} refers to the context of a word within text, this same quote applies to geography. Another justification that resembles TFL comes from \cite{firth1957synopsis}, stating "You shall know a word by the company it keeps!" 

Words and territories are also similar since one represents them as categorical variables with a large cardinality. For this reason, it isn't surprising that the classical models for NLP tasks and geographic prediction are similar. Simple representations of text like bag-of-words and n-grams are similar to simple representations of territories like one-hot encodings. Further techniques use smoothing to account for unknown words or territories, while more sophisticated methods use graphical models like hidden Markov models or conditional autoregressive models to account for textual or geographic autocorrelation. Most recent models in NLP use word embeddings \cite{mikolov2017advances}, and researchers are starting to study geographic embeddings. Table \ref{tab:nlp-vs-spatial} summarizes the comparisons outlined in this paragraph. 

\begin{table}[h]
	\centering\setlength\tabcolsep{3.5pt}
	\begin{tabular}{|c|c|c|}
		\hline
		                              & NLP                                                                                    & Geography                                                                                                \\ \hline
		Fondamental justification     & \thead{Distributional hypothesis \\\cite{harris1954distributional, firth1957synopsis}} & \thead{Tobler's first law of geography \\\cite{tobler1970computer} }                                     \\ \hline
		Classical tools \& techniques & \thead{Bag-of-words \& n-grams\\n-gram smoothing (Laplace)\\Hidden Markov models}      & \thead{One-hot encoding of territories\\Smoothing (kriging, splines)\\Conditional autoregressive models} \\ \hline
		Representation techniques     & \thead{Word embeddings \\ \cite{mikolov2013efficient}\\\cite{devlin2019bert}}          & \thead{Geographic embeddings \\ \cite{fu2019efficient}, \\\cite{blier2020encoding}}                      \\ \hline
	\end{tabular}
	\caption{Comparing contributions and techniques for NLP and geographic models}\label{tab:nlp-vs-spatial}
\end{table}

Attribute \ref{att:coordinate2} proposes to use coordinates instead of polygons to construct geographic embeddings. This is analogous to using character-level embeddings instead of word embeddings. NLP researchers use character-level embeddings to understand the semantics of an unknown word (a word which we never saw in a training corpus). 

The principal difference between words and territories is that territories are not interchangeable or synonymous. Instead, we capture the co-occurrence of relevant geographic information, see attribute \ref{att:external3}.

    % !TeX spellcheck = en_US

\section{Convolution-based geographic embedding model}\label{sec:crae}

In this section, we describe an approach to construct geographic embeddings. We will explain the representation model choices for the encoder in \ref{item:representation1}, the decoder in \ref{item:representation2} and the evaluation in \ref{item:representation4}. For the embeddings to have all desirable attributes of geographic embeddings enumerated in the previous section, we must modify the process to account for geographic data, adding a data preparation step. Our focus is on constructing a geographic representation model, so we omit the (optional) \ref{item:representation3} of combining representations with other sources of data in this section. 

\subsection{Preparing the data}\label{ss:prep}

Suppose we are interested in creating a geographic feature that characterizes a location \textbf{s}. Following attribute \ref{att:external3} (external), we must first collect geographic information for location \textbf{s}. Objects characterized by their location in space can be stored as point coordinates or polygons (boundary coordinates) \cite{anselin2010geoda}. Point coordinate data corresponds to the measurements of a variable at an individual location. It can indicate the presence of a phenomenon (set of locations where accidents occurred) or the measurement of a variable at a particular location (age of a building at a location) \cite{diggle2013statistical}. Polygon data aggregates variables from all point patterns in a territory. 
%Since public data is usually aggregated for privacy reasons; most external geo-localized data is published as polygon data.
In Figure \ref{fig:dist-points-2}, the individual marks A, B, C and D are point patterns, and the different shaded areas are polygons. 

To construct the representation model, we assume that we have one or many external datasets in polygon format, with a vector of data for each polygon. This is the case for census data in North America. Suppose the coordinate of \textbf{s} is located within the boundaries of polygon $\textbf{S}$, then one associates the external geographic data from polygon \textbf{S} to location \textbf{s}. We will use the notation $\boldsymbol{\gamma} \in \mathbb{R}^d$ to denote the vector of geographic variables.  

In usual (non-spatial) situations, the type and format of data determines the topology of the encoder. Geographic information is often vectorial (called spatial explanatory variables \cite{diggle2007model}) so that the associated neural network encoder would be a fully-connected neural network. A fully-connected neural network encoder would satisfy the \textit{external} attribute, but not \textit{TFL} or \textit{coordinate}. To understand why the \textit{coordinate} is not entirely satisfied, we consider one external data source from one set of polygons. In this case, observations within the same polygon will have the same geographic variables $\boldsymbol{\gamma}$, so the geographic data remains polygon-based. Instead, we modify the input data for the representation model to satisfy the \textit{TFL} and \textit{coordinate} attributes. To create an embedding of location \textbf{s}, we will not only use information from location \textbf{s}, but also data from surrounding locations. Since each coordinate within a polygon may have different surrounding locations, the resulting embeddings will not be polygon-based, satisfying the \textit{coordinate} attribute. 

An approach to satisfy \textit{TFL} is to use an encoder that includes a notion of nearness. Convolutional neural networks (CNNs) have the property of sparse interactions, meaning the outputs of a convolutional operation are features that depend on local subsets of the input \cite{goodfellow2016deep}. CNNs accept matrix data as input. For this reason, we will define our neighbors to form a grid
%\footnote{What we call a grid is related to a raster, a geographic data format.} 
around a central location. The representation model's input data is the geographic data square cuboid from \cite{blier2020encoding}, which we present in Algorithm \ref{algo:generate_image} and explain below. One notices that the geographic data square cuboid is a specific case of a multiband raster, datacube or multidimensional array. 

\begin{algorithm}
	\KwIn{Center coordinate $(lon, lat)$, width $w$, size $p$, geographic data sources}
	\KwOut{Geographic data square cuboid}
	\nl generate empty grid of dimension $p \times p$ and width $1$ with the set
	$$\bigcup_{k = 0}^{p-1}\bigcup_{j = 0}^{p-1} \left[\frac{1}{2}\left(2k - p + 1 \right), ~\frac{1}{2}\left(2j - p + 1 \right)  \right],$$
	store the coordinates of this set of points in a matrix $\delta_0$ (each column is a coordinate, the first row is the longitude, the second row is the latitude)\;
	\nl scale (multiply) by $w$\;
	\nl sample $\theta \sim Unif(0, 360)$ and rotate the scaled matrix, 
	$ \displaystyle R(w\delta_0) = w \left(\begin{array}{cc}
	\cos \theta  & - \sin \theta\\ \sin \theta & \cos \theta
	\end{array}
	\right) \delta_0$\;
	\nl translate matrix by $(lon, lat)'$, to get $\displaystyle \delta = R(w \delta_0) + \left(\begin{array}{c}
	lon\\ lat
	\end{array}\right)$\;
	\nl \ForEach{external geographic data source}{
		\nl \ForEach{coordinate in $\delta$}{
			\nl allocate coordinate to corresponding polygon index\;
			\nl append vector data to current coordinate
		}
	}
	\caption{Generating a geographic data square cuboid} \label{algo:generate_image}
\end{algorithm}

\begin{figure}[ht]
	\centering
	\begin{subfigure}[b]{0.24\textwidth}
		\centering
		\resizebox{0.75\textwidth}{!}{\begin{tikzpicture}	
				
				\draw[thick,color=gray,step=.5cm,
				dashed] (7, 7) grid (7, 7);
				\draw[->] (-7,0) -- (7,0)
				node[below right] {\LARGE };
				\draw[->] (0,-7) -- (0,7)
				node[left] {\LARGE };

				\tstar{0.05}{0.5}{8}{0}{thick,fill=red}
				\foreach \n in {-3.5, -2.5, ..., 3.5}
				\foreach \j in {-3.5, -2.5, ..., 3.5}
				\node at (\n * 0.7, \j * 0.7)[circle,fill,inner sep=3pt, color = red]{};
				
				\draw[thick] (-3.5 * 0.7, -3.5 * 0.7)--(-2.5 * 0.7, -3.5 * 0.7) node[pos=0.6, inner sep=0pt, outer sep=0pt, below = 0.5cm](x){\Huge 1};
				
%				\node at (0.75, 0) {$(0, 0)$};
				
		\end{tikzpicture}}
		\caption{Unit grid}
		%		\label{fig:}
	\end{subfigure}
	\hfill
	\begin{subfigure}[b]{0.24\textwidth}
		\centering
		\resizebox{0.75\textwidth}{!}{\begin{tikzpicture}
				
				\draw[thick,color=gray,step=.5cm,
				dashed] (7, 7) grid (7, 7);
				\draw[->] (-7,0) -- (7,0)
				node[below right] {\LARGE };
				\draw[->] (0,-7) -- (0,7)
				node[left] {\LARGE };

				\tstar{0.05}{0.5}{8}{0}{thick,fill=red}
				\foreach \n in {-3.5, -2.5, ..., 3.5}
				\foreach \j in {-3.5, -2.5, ..., 3.5}
				\node at (\n, \j)[circle,fill,inner sep=3pt, color = red]{};
				
				\draw[thick] (-3.5, -3.5)--(-2.5, -3.5) node[pos=0.6, inner sep=0pt, outer sep=0pt, below = 0.5cm](x){\Huge w};
				
%				\node at (0.75, 0) {$(0, 0)$};
		\end{tikzpicture}}
		\caption{Scale}
		%		\label{fig:}
	\end{subfigure}
	\begin{subfigure}[b]{0.24\textwidth}
		\centering
		\resizebox{0.75\textwidth}{!}{\begin{tikzpicture}
				
				%		\draw[thick, step=1cm] (-4,-4) grid (4,4);
				
				\draw[thick,color=gray,step=.5cm,
				dashed] (7, 7) grid (7, 7);
				\draw[->] (-7,0) -- (7,0)
				node[below right] {\LARGE };
				\draw[->] (0,-7) -- (0,7)
				node[left] {\LARGE };
				
				\tstar{0.05}{0.5}{8}{0}{thick,fill=red}
				\foreach \n in {-3.5, -2.5, ..., 3.5}
				\foreach \j in {-3.5, -2.5, ..., 3.5}
				%		\node at (\n, \j)[circle,fill,inner sep=3pt, color = red]{};
				\node at (\n * 0.525322 - \j * 0.8509035, \n * 0.8509035 + \j * 0.525322)[circle,fill,inner sep=3pt, color = red]{};
				
				\draw[thick] (1.139535, -4.816789)--(1.664857, -3.965886) node[pos=0.6, inner sep=0pt, outer sep=0pt, below = 0.5cm, rotate=60](x){\Huge w};
				
%				\node at (0.75, 0.5) {$(0, 0)$};

				\draw[->,semithick] (45:6.363961) arc[radius=6.363961, start angle=45, end angle=110];

		\end{tikzpicture}}
		\caption{Rotate}
		%		\label{fig:}
	\end{subfigure}\hfill
	\begin{subfigure}[b]{0.24\textwidth}
		\centering
		\resizebox{0.75\textwidth}{!}{\begin{tikzpicture}
				
				\draw[thick,color=gray,step=.5cm,
				dashed] (7, 7) grid (7, 7);
				\draw[->] (-7,0) -- (7,0)
				node[below right] {\LARGE };
				\draw[->] (0,-7) -- (0,7)
				node[left] {\LARGE };
				
				\tstar{0.05}{0.5}{8}{0}{thick,fill=red, xshift=2cm, yshift=2cm}
				\foreach \n in {-3.5, -2.5, ..., 3.5}
				\foreach \j in {-3.5, -2.5, ..., 3.5}
				%		\node at (\n, \j)[circle,fill,inner sep=3pt, color = red]{};
				\node at (\n * 0.525322 - \j * 0.8509035 + 2, \n * 0.8509035 + \j * 0.525322 + 2)[circle,fill,inner sep=3pt, color = red]{};
				
				\draw[thick] (1.139535 +2, -4.816789 +2)--(1.664857 +2, -3.965886 +2) node[pos=0.6, inner sep=0pt, outer sep=0pt, below = 0.5cm, rotate = 60](x){\Huge w};
%				\node at (0.75, 0.5) {$(lon, lat)$};
		\end{tikzpicture}}
		\caption{Translate}
		%		\label{fig:}
	\end{subfigure}
	\caption{Creating the grid of neighbors}\label{fig:neighbors}
\end{figure}
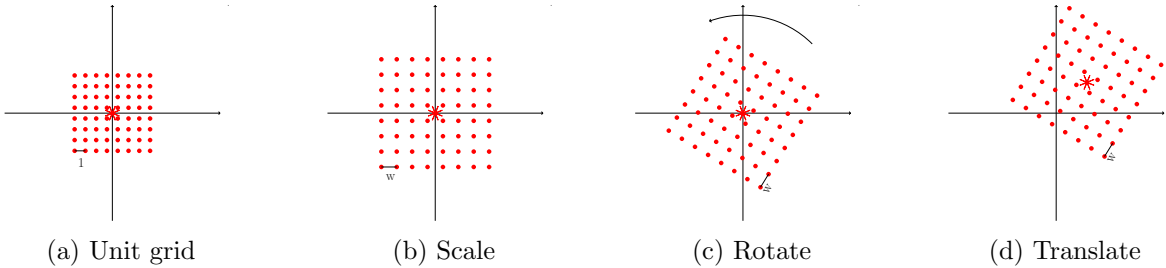

To create the geographic data cuboid for a location \textbf{s} with coordinate $(lon, lat)$, we span a grid of $p\times p$ neighbors, centered around $(lon, lat)$. Each neighbor in the grid is separated horizontally and vertically by a distance of $w$. The steps 1 to 4 of Algorithm \ref{algo:generate_image} create the matrix of coordinates for the neighbors, and we illustrate each transformation in Figure \ref{fig:neighbors}.

\begin{figure}[ht]
	\minipage{0.45\textwidth}
	\includegraphics[width=\linewidth]{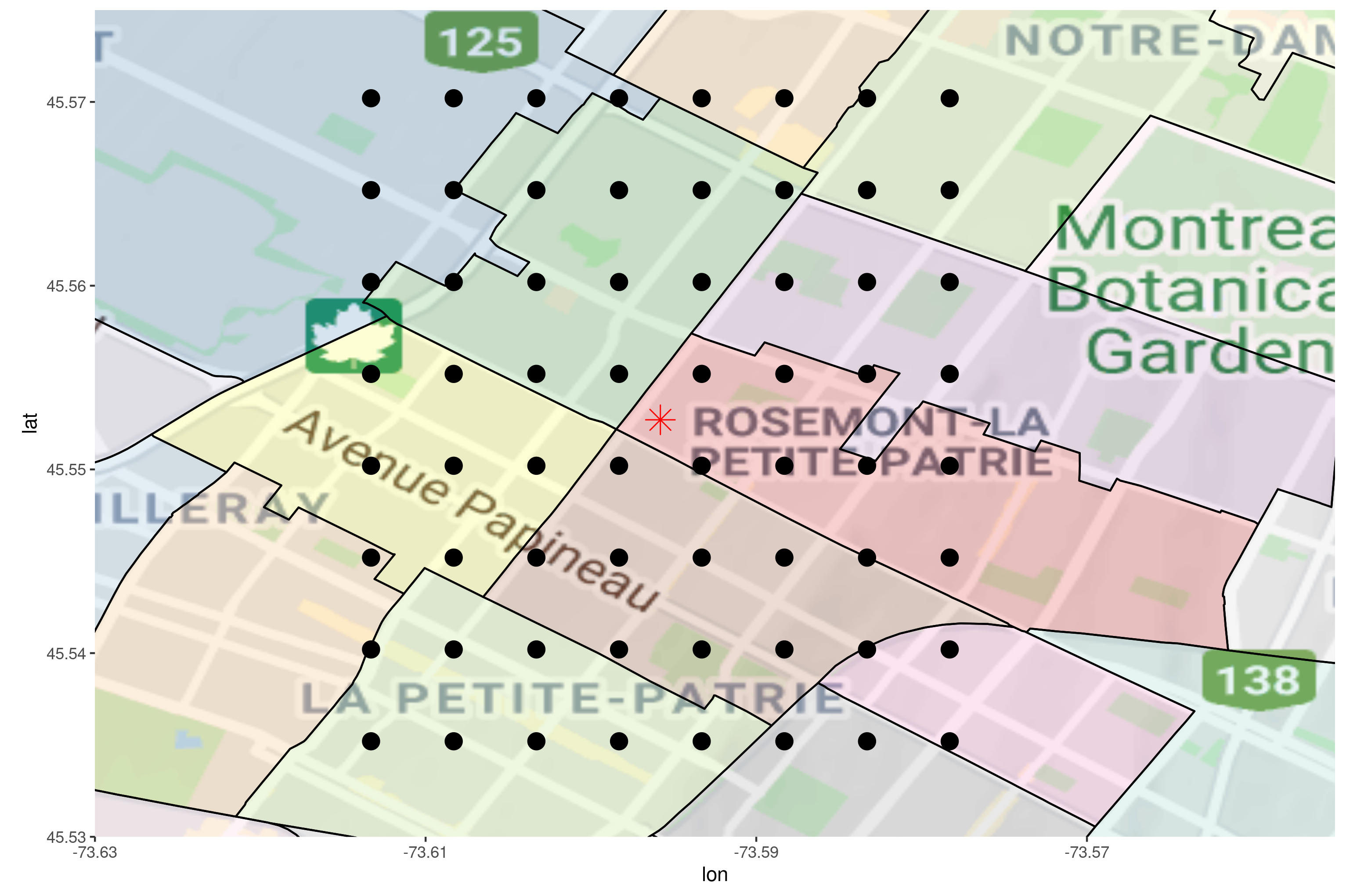}
	\caption{Set of neighbor coordinates $\delta$}\label{fig:span}
	\endminipage\hfill
	\minipage{0.45\textwidth}
	\resizebox{0.72\textwidth}{0.62\textwidth}{\begin{tikzpicture}
	
	\draw[thick, step=1cm] (-4,-4) grid (4,4);
	\tstar{0.05}{0.5}{8}{0}{thick,fill=red}
	\foreach \n in {-3.5, -2.5, ..., 3.5}
	\foreach \j in {-3.5, -2.5, ..., 3.5}
	\node at (\n, \j)[circle,fill,inner sep=3pt, color = red]{};
	
	\draw[thick] (-4, 4) -- (-12,12);
	\draw[thick] (-4, -4) -- (-12,4);
	\draw[thick] (-12,12) -- (-12,4);
	\draw[thick] (-12,12) -- (-4,12);
	\draw[thick] (-4,12) -- (4, 4);
	
	%	\draw[thick] (-4.5,4.5) -- (3.5, 4.5);
	%	\draw[thick] (-4.5,4.5) -- (-4.5, -3.5);	
	
	\foreach \n in {1, 2, 3, 6, 7}	
	\draw[thick] (-4 - \n, 4 + \n) -- (4 - \n, 4 + \n);
	\foreach \n in {1, 2, 3, 6, 7}	
	\draw[thick] (-4 - \n,4 + \n) -- (-4-\n, -4 + \n);	
	
	\foreach \n in {1, 2, 3}	
	\node at (0 - \n * 1 + 0.5, 3.5 + \n * 1) {\Huge Variable \n};
	
	\node at (-11.25 + 3.5, 11.5) {\Huge Variable $d$};
	\node at (-11.25 + 3.5 + 1, 11.5 - 1) {\Huge Variable $d$-1};
	
	\draw[loosely dotted, very thick] (-4.5,8.5) -- (-4.2,8.2);
	\draw[loosely dotted, very thick] (-8.5,4.5) -- (-8.2,4.2);
	
	\end{tikzpicture}}
	
	\caption{Data square cuboid $\underline{\boldsymbol{\gamma}}_\delta$}\label{fig:grid}
	\endminipage\hfill
\end{figure}

Each coordinate in $\delta$ has geographic variables $\boldsymbol{\gamma} \in \mathbb{R}^d$, extracted from different external data sources of polygon data. The set of variables for every location in $\delta$, which we note $\underline{\boldsymbol{\gamma}}_\delta$, forms a square cuboid of dimension $p\times p \times d$, which we illustrate in Figure \ref{fig:grid}. We can interpret the geographic data square cuboid as an image with $d$ channels, with one channel for each variable. Algorithm \ref{algo:generate_image} presents the steps to create the data square cuboid for a single location. Lines 1 to 4 generate the set $\delta$ for the center location (the matrix of neighbor coordinates), while lines 4 to 7 fill each coordinate's variables. The random rotation is present such that our model does not exclusively learn effects in the same orientation. 

If the grid size $p$ is even, $\delta$ does not include the center coordinate $(lon, lat)$. This means the geographic data square cuboid depends only on neighbor information and not information of the center point itself.

\subsection{Constructing the encoder}\label{ss:encoder}

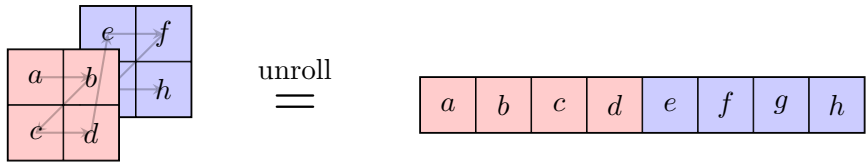
\begin{figure}[ht]
	\centering
	\begin{tikzpicture}
	
	\matrix[matrix of math nodes, draw,thick,inner sep=0pt, cells={nodes={minimum width=1.9em,minimum height=1.9em,
			draw,very thin,anchor=center,fill=blue!20}}, xshift=2.5em,yshift=1.5em](efgh){
		e& f\\g&h\\
	};
	
	\draw [arrow, opacity=0.25] (efgh-1-1.center) -- (efgh-1-2.center);
	\draw [arrow, opacity=0.25] (efgh-1-2.center) -- (efgh-2-1.center);
	\draw [arrow, opacity=0.25] (efgh-2-1.center) -- (efgh-2-2.center);

	\matrix[matrix of math nodes, draw,thick,inner sep=0pt, cells={nodes={minimum width=1.9em,minimum height=1.9em,
			draw,very thin,anchor=center,fill=red!20}}, xshift=0em,yshift=0em](abcd){
		a&b\\c&d\\
	};
	
	\draw [arrow, opacity=0.25] (abcd-1-1.center) -- (abcd-1-2.center);
	\draw [arrow, opacity=0.25] (abcd-1-2.center) -- (abcd-2-1.center);
	\draw [arrow, opacity=0.25] (abcd-2-1.center) -- (abcd-2-2.center);
	\draw [arrow, opacity=0.25] (abcd-2-2.center) -- (efgh-1-1.center);

	\node[xshift = 8em, scale=2, yshift = 0em] (eq) {=};
	\node[xshift = 8em, scale=1, yshift = 1.25em] (unroll) {unroll};
	
	\matrix[matrix of math nodes, draw,thick,inner sep=0pt, cells={nodes={minimum width=1.9em,minimum height=1.9em,
			draw,very thin,anchor=center}}, xshift=20em,yshift=0em]{
		|[fill=red!20]|a&|[fill=red!20]|b&|[fill=red!20]|c&|[fill=red!20]|d&|[fill=blue!20]|e&|[fill=blue!20]|f&|[fill=blue!20]|g&|[fill=blue!20]|h\\
	};

\end{tikzpicture}
	\caption{Unrolling example for a $2\times 2 \times 2$ cuboid. Different colors are different variables}\label{fig:unroll}
\end{figure}

As explained in Section \ref{sec:representations}, the encoder generates a latent representation from a bottleneck, which we then extract as embeddings. Above, we constructed the input data to have a square cuboid shape to use convolutional neural networks as an encoder. The encoder has convolutional operations, reducing the size of the hidden square cuboids (also called feature maps in computer vision) after each set of convolutions. Then, we \textit{unroll} (from left to right, top to bottom, front to back, see Figure \ref{fig:unroll} for an illustration) the square cuboid into a vector. The unrolled vector will be highly autocorrelated: since the output of the convolutions includes local features, features from the same convolutional layer will be similar to each other, causing collinearity if we use the unrolled vector directly within a GLM. For this reason, we place fully-connected layers after the unrolled vector. The last fully-connected layer of the encoder is the geographic embedding vector, noted $\boldsymbol{\gamma}^*$. This layer typically has the smallest dimension of the entire neural network representation model. For a geographic embedding of dimension $\ell$, then the encoder will be a function $\mathbb{R}^{p\times p \times d} \to \mathbb{R}^\ell$.

We have now constructed the encoder for our representation model. This encoder satisfies desirable attributes \ref{att:tfl1} to \ref{att:external3}  since
\begin{enumerate}
	\item the encoder is a convolutional neural network, which captures sparse interactions from neighboring information, encoding the notion of nearness;
	\item the embeddings are coordinate-based: as the center coordinate moves, the neighboring coordinates also move, so center coordinates within the same polygon may have different geographic data square cuboids;
	\item the model uses external data as opposed to one-hot encodings of territories.
\end{enumerate}

\subsection{Constructing the decoder}\label{ss:decoder}

The decoder guides the type of information that the embeddings encode, i.e., selects which domain knowledge to induce into the representations, see Section \ref{sec:representations} for details on the decoding procedure. If one has response variables related to the insurance domain for the whole dataset (i.e., for all coordinates within the territory for which we construct embeddings), we could train the decoder with transfer learning, such that the representation model learns important information related to insurance. In our case, the external dataset that covers Canada, and we do not have insurance statistics over the entire country. For this reason, our decoder will attempt to predict itself, i.e., the input variables $\boldsymbol{\gamma}$. 

The original model for geographic embeddings we explored was the convolutional regional autoencoder (CRAE), presented in \cite{blier2020encoding}. The input of CRAE is the geographic data cuboid, and the output is also the geographic data cuboid. The neural network architecture is called a convolutional autoencoder since the model's objective is to reconstruct the input data after going through a bottleneck of layers. The decoder in CRAE is a function $\mathbb{R}^\ell \to \mathbb{R}^{p\times p \times d}$. One can interpret CRAE as \textit{using contextual variables to predict the same contextual variables}. The loss function for CRAE is the average reconstruction error. For a dataset of $N$ coordinates, the loss function is 
\begin{equation}\label{eq:reconstruct-crae}
	\mathcal{L} = \frac{1}{N}\sum_{i = 1}^{N}\left|\left|g\left(f\left(\underline{\boldsymbol{\gamma}}_{\delta_i}\right)\right) - \underline{\boldsymbol{\gamma}}_{\delta_i}\right|\right|^2, %+ \lambda \left\lvert\left\lvert W \right\rvert\right\rvert^2,
\end{equation}
where $f$ is the encoder,
%\footnote{Notice that $\boldsymbol{\gamma}^* = f\left(\underline{\boldsymbol{\gamma}}_{\delta_i}\right)$.} 
$g$ is the decoder, $\underline{\boldsymbol{\gamma}}_{\delta_i} \in \mathbb{R}^{p\times p \times d}$ is the geographic data cuboid for the coordinate of location $i$ and $||\cdot||$ is the euclidean norm.
%\footnote{Loss functions sometimes have penalty terms for the size of weights. As explained in Section \ref{sec:implementation}, the neural networks do not overfit, so regularization is not necessary.} 
However, attribute \ref{att:coordinate2} only requires information from the grid's central location, not the entire grid. Therefore, much of the information contained within CRAE embeddings captures irrelevant or redundant information. 

\begin{figure}[ht]
	\centering
	\resizebox{0.75\textwidth}{!}{
	\begin{tikzpicture}
	\tikzstyle{connection}=[ultra thick,every node/.style={sloped,allow upside down},draw=\edgecolor,opacity=0.7]
	\tikzstyle{copyconnection}=[ultra thick,every node/.style={sloped,allow upside down},draw={rgb:blue,4;red,1;green,1;black,3},opacity=0.7]
	
%	\path[use as bounding box] (0, -2.5) rectangle (20, 4);

	\pic[shift={(0,0,0)}] at (0,0,0) 
	{Box={
			name=conv1,
			caption=,
			xlabel={{, }},
			zlabel=\Large ~~Input,
			fill=green,
			opacity=0.25,
			height=16,
			width=8,
			depth=16
		}
	};
	
	\pic[shift={(1.5,0,0)}] at (conv1-east) 
	{Box={
			name=conv2,
			caption= ,
			xlabel={{, }},
			zlabel=,
			fill=red,
			opacity=0.25,
			height=8,
			width=16,
			depth=8
		}
	};
	
	\pic[shift={(1.5,0,0)}] at (conv2-east) 
	{Box={
			name=conv3,
			caption=,
			xlabel={{, }},
			zlabel=,
			fill=red,
			opacity=0.25,
			height=4,
			width=16,
			depth=4
		}
	};

	\pic[shift={(1.5,0,0)}] at (conv3-east) 
	{Box={
			name=fc1,
			caption=,
			xlabel={{, }},
			zlabel=\Large ~~ ~ ~~~~Unroll,
			fill=red,
			opacity=0.25,
			height=1,
			width=1,
			depth=48
		}
	};
	
	\pic[shift={(1.5,0,0)}] at (fc1-east) 
	{Box={
			name=fc2,
			caption= ,
			xlabel={{, }},
			zlabel=\Large ,
			fill=red,
			opacity=0.25,
			height=1,
			width=1,
			depth=16
		}
	};
	
	\pic[shift={(1.5,0,0)}] at (fc2-east) 
	{Box={
			name=soft1,
			caption=,
			xlabel={{,}},
			zlabel=\Large Embedding,
			fill=violet,
			opacity=0.3,
			height=1,
			width=1,
			depth=8
		}
	};
	
	\pic[shift={(1.5,0,0)}] at (soft1-east) 
	{Box={
			name=fc3,
			caption=,
			xlabel={{,}},
			zlabel=\Large ,
			fill=blue,
			opacity=0.25,
			height=1,
			width=1,
			depth=16
		}
	};
	
	\pic[shift={(1.5,0,0)}] at (fc3-east) 
	{Box={
			name=fc4,
			caption=,
			xlabel={{,}},
			zlabel=\Large ~~~~Output,
			fill=blue,
			opacity=0.25,
			height=1,
			width=1,
			depth=32
		}
	};

	\draw[->, shorten <=2pt,shorten >=2pt, thick] (conv1-east) -- (conv2-west);
	\draw[->, shorten <=2pt,shorten >=2pt, thick] (conv2-east) -- (conv3-west);
	\draw[->, shorten <=2pt,shorten >=8pt, thick] (conv3-east) -- (fc1-west);
	\draw[->, shorten <=8pt,shorten >=8pt, thick] (fc1-east) -- (fc2-west);
	\draw[->, shorten <=8pt,shorten >=8pt, thick] (fc2-east) -- (soft1-west);
	\draw[->, shorten <=8pt,shorten >=8pt, thick] (soft1-east) -- (fc3-west);
	\draw[->, shorten <=8pt,shorten >=8pt, thick] (fc3-east) -- (fc4-west);

;	\end{tikzpicture}}
	\caption{Convolution-based geographic embedding model
%		\todo[inline, color=cyan!40]{E: la figure n'est pas claire. Est-ce que cette image permet au lecteur de dire que l'auteur veut servir un bol de ceréales avec du lait? Miam.}
	}\label{fig:csem}
\end{figure}
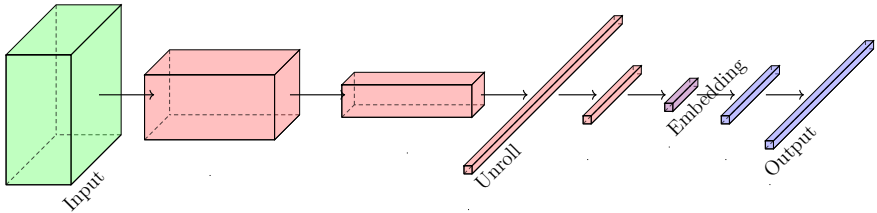

In this paper, we improve CRAE by changing the decoder. One can interpret the new model as using \textit{contextual variables to predict the central variables}, directly satisfying the \textit{coordinate} attribute. This same motivation was suggested for natural language processing in \cite{collobert2008unified} and applied in a model called Continuous Bag of Words (CBOW) \cite{mikolov2013efficient}. Instead of reconstructing the entire geographic data cube, the decoder attempts to predict the variables $\boldsymbol{\gamma}$ for location \textbf{s}. Therefore, the decoder is a series of fully-connected layers that act as a function $\mathbb{R}^\ell \to \mathbb{R}^{d}$, where $\ell \ll d$. The loss function for CBOW-CRAE is also the average reconstruction error, but on the vector of variables for the central location instead on the entire geographic data square cuboid: 
\begin{equation}\label{eq:reconstruct-cbow-crae}
\mathcal{L} = \frac{1}{N}\sum_{i = 1}^{N}\left|\left|g\left(f\left(\underline{\boldsymbol{\gamma}}_{\delta_i}\right)\right) - \boldsymbol{\gamma}_i\right|\right|^2.
\end{equation}
Figure \ref{fig:csem} illustrates the entire convolution-based model architecture of CBOW-CRAE.

\subsection{Evaluating representations}

There are two types of evaluations for embeddings: the most important are extrinsic evaluations, but intrinsic evaluations are also significant \cite{jurafsky2000speech}. One evaluates embeddings extrinsically by using them in downstream tasks and comparing their predictive performance. In P\&C actuarial ratemaking, one can use the embeddings within ratemaking models for different lines of business and select the embedding model that minimizes the loss on out-of-sample data. 

According to our knowledge, there are no intrinsic evaluation methods for geographic embeddings. We propose one method in this paper and discuss other approaches in the conclusion. Intrinsic evaluations determine if the embeddings behave as expected. To evaluate embeddings intrinsically, one can determine if they possess the three attributes proposed in Section \ref{sec:spatial-rep}. Due to the geographic data cuboid construction, geographic embeddings already satisfy attributes \textit{coordinate} and \textit{external}, so these attributes do not need intrinsic evaluations. To validate \textit{TFL}, we can plot individual dimensions of the embedding vector on a map. Embeddings that satisfy \emph{TFL} vary smoothly, and one can inspect this property visually. Section \ref{ss:implicit} presents the implicit evaluation for the implementation on Canadian census data. 

    % !TeX spellcheck = en_US
\section{Implementations of geographic embeddings}\label{sec:implementation}

In this section, we present an implementation of geographic embeddings using census data in Canada. We select census data since they contain crucial socio-demographic information on the customer's geography; geographic embeddings trained with census data will compress this information. One could also use natural (ecosystem, landform, weather) information to create geographic embeddings that capture natural geographic risk. We first present the census data in Canada, along with issues related to some of its characteristics. Then, we explain the architectural choices and implementations for the geographic data square cuboid, the encoder and the decoder. Finally, we perform intrinsic evaluations of the geographic embeddings. 

Our strategy for constructing geographic embeddings contains two types of hyperparameters: general neural network hyperparameters, and specific hyperparameters related to the particular model architecture. \textit{General hyperparameters} control the method used to optimize the neural network parameters, especially the optimization routine: these include the choice of the optimizer, the batch size, the learning rate, the weight decay, the number of epochs and the patience for the learning rate reduction. We select general hyperparameters using grid-search, along with personal experience with neural networks.  We refer the reader to other texts such as \cite{goodfellow2016deep} for general tips on hyperparameter search. We will limit our discussion of general hyperparameter search to mentioning the optimal parameters that we use. \textit{Specific hyperparameters} are associated with the specific neural network architecture and includes the square cuboid width $w$ and the pixel size $p$ to generate a geographic data square cuboid. Other specific hyperparameters determine the neural network architecture, including the shape and depth of convolutions and fully connected layers. Tuning these hyperparameters requires experience with neural networks since the combinations of specific hyperparameters are much too large to determine using grid-search. We will explain our thought process: starting with accepted heuristics, then manually exploring the hyperparameter space to determine the optimal architecture. 

\subsection{Canadian census data}

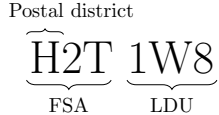
\begin{figure}
	\centering
	\resizebox{3cm}{!}{
		\begin{tikzpicture}
		\node at (0,0) {\Huge H2T 1W8};
		\draw [decorate,decoration={brace,amplitude=5pt,mirror, raise = 2ex}] (-1.8,-0.1) -- (-0.2, -0.1) node[midway,yshift=-0.8cm]{FSA};
		\draw [decorate,decoration={brace,amplitude=5pt,mirror, raise = 2ex}] (0.1,-0.1) -- (1.8, -0.1) node[midway,yshift=-0.8cm]{LDU};
		\draw [decorate,decoration={brace,amplitude=3pt, raise = 2ex}] (-1.8,0.1) -- (-1.1, 0.1) node[midway,yshift=0.8cm, xshift = 0.5cm]{Postal district};
		\end{tikzpicture}}
	\caption{Deconstruction of a Canadian postal code}
	\label{fig:FSA}
\end{figure}

We now present some characteristics of Canadian census data. Statistics Canada publishes this data every five years, and our implementation uses the most recent version (2016). The data is aggregated into polygons called forward sortation areas, which correspond to the first three characters of a Canadian postal code; see Figure \ref{fig:FSA} for the decomposition of a postal code. {Statistics Canada aggregates the public release of census data to avoid revealing confidential and individual information. The data is also available at the dissemination area polygon level, which is more granular than FSA. We work with FSAs because they are simpler to explain.} There are 1 640 FSAs in Canada, and each polygon in Figure \ref{fig:span} represents a different FSA. The grid of neighbor coordinates of Figure \ref{fig:span} contains 8 points from the same FSA as the central location, represented by the red star.

The first issue with using census data for insurance pricing is the use of protected attributes, i.e., variables that should legally or ethically not influence the model prediction. One example is race and ethnicity \cite{frees2015analytics}. Territories may exhibit a high correlation with protected attributes like ethnic origin. To construct the geographic embeddings, we discard all variables related to ethnic origin (country of origin, mother tongue, citizen status). We retrain only variables that a Canadian insurance company could use for ratemaking. In Appendix \ref{sec:census-detail}, we provide a complete list of the categories of variables within the census dataset, and we denote with an asterisk the categories of variables that we omit. What remains is information about age, education, commute, income and others, and comprises 512 variables that we denote $\gamma$. Removing protected attributes from a model is a technique called anti-classification \cite{corbett-davies2018measure}, or fairness through unawareness \cite{kusner2018counterfactual}, which does not eliminate discrimination entirely and in some cases may increase it \cite{kusner2018counterfactual}. Studying discrimination-free methods to construct geographic embeddings is kept as future work. For analysis and discussion of discrimination in actuarial ratemaking, see \cite{lindholm2020discrimination}. 

It is common practice in machine learning to normalize input variables such that they are all on the same scale. Two reasons are that stochastic gradient-based algorithms typically converge faster with normalized variables, and un-normalized variables have different impacts on the regularization terms of the loss function \cite{shalev2014understanding}. For autoencoders, an additional reason is that the model's output is the reconstruction of many variables, so variables with a higher magnitude will generate a disproportionately large reconstruction error relative to other variables (so the model will place more importance on reconstructing these variables). Normalization requires special attention for aggregated variables like averages, medians, and sums: some must be aggregated with respect to a small number of observations, others with respect to the population within the FSA of interest and others with the Canadian average. For example, the FSA population is min-max normalized (see \cite{han2011data}) with respect to all FSAs in Canada. For age group proportions (for instance, the proportion of 15 to 19-year-olds), we normalize with respect to the current FSA's population size.  

When using Algorithm \ref{algo:generate_image}, some coordinates do not belong to a polygon index, which happens when the coordinate is located in a body of water. To deal with this situation, we created a vector of missing values filled with the value 0. 

Finally, we project all coordinates to Atlas Lambert \cite{lambert1772beitrage, snyder1987map} to reduce errors when computing distances. Due to Earth's curvature, computing Euclidean distance with degree-based coordinates like GPS underestimates distances; see \cite{torge2012geodesy} for illustrations. When changing the coordinate system to Atlas Lambert, one can compute distances using traditional euclidean metrics with greater accuracy. 

\subsection{Geographic data square cuboid}

Now that we have prepared the census data, we can construct the geographic data cuboid for our implementation. The parameters for the geographic data square cuboid include square width $w$ and pixel size $p$. One can interpret these values as smoothing parameters. The square width $w$ affects the geographic scale of the neighbors, while the pixel size $p$ determines the density of the neighbors. For very flexible embeddings, one can choose small $w$ so that the geographic embeddings will capture local geographic effects. If the embeddings are too noisy, then one can increase $p$ to stabilize the embedding values. Ultimately, the importance is the span of the grid, determined by the farthest sampled neighbors. For an even value of $w$, the closest sampled neighbor coordinates are the four corners surrounding the central location at a distance of $p/\sqrt{2}$ units from the central location. The farthest sampled neighbors are the four outermost corners of the grid at a distance of $p(w-1)/\sqrt{2}$ units. Selecting the best parameters is a tradeoff between capturing local variations in densely populated areas and smoothing random fluctuations in rural areas. Insurance companies could construct embeddings for rural portfolios and urban portfolios with different spans to overcome this compromise. Another solution consists of letting the parameters depend on local characteristics of the population, such as population density (select parameters such that the population contained within a grid is above a threshold) or the range of a variogram (select parameters such that the observations within the grid still exhibit geographic autocorrelation). 

We consider square widths of $w= \{8, 16\}$ and pixel sizes of $p = \{50, 100, 500\}$ meters. Our experiments showed that a square width of $w=16$ and a pixel size of $p = 100$ meters produced smaller reconstruction errors. The geographic data square cuboid then samples $16^2 = 256$ neighbors, the closest one being at a distance of 71 meters and the farthest one at 1061 meters from the center location. Since we have 512 features available for every neighbor, the size of the input data is $16 \times 16 \times 512$ and contains 131 072 values. 

The dataset used to train the representation models is composed of a geographic data square cuboid for every postal code in Canada (888 533 by the time of our experiments). A dataset containing the values of every geographic data square cuboid would take 2To of memory, too large to hold in most personal computers. For this reason, we could not generate the complete dataset. However, one can still train geographic embeddings on a personal computer by leveraging the fact that many values in the dataset repeat. For each postal code, one generates the grid of neighbor coordinates (steps 1 to 4 of Algorithm \ref{algo:generate_image}) and identifies the corresponding FSA for each coordinate (step 6 of Algorithm \ref{algo:generate_image}). We then unroll the grid from left to right and from top to bottom into a vector of size 256. For memory reasons, we create a numeric index for every FSA (A0A = 1, A0B = 2, $\dots$, Y1A = 1 640), and store the list of index for every postal code in a CSV file (1GB). Another dataset contains the normalized geographic variables $\boldsymbol{\gamma}$ for every FSA (15.1MB). Most modern computers can load both datasets in RAM. During training, our implementation retrieves the vector from the index and retrieves the values associated with each FSA (step 8 of Algorithm \ref{algo:generate_image}), then rolls the data into geographic data square cuboids. 

\subsection{Encoders}

This subsection will detail specific architecture choices for encoders with Canadian census data, presenting two strategies to determine the optimal architecture. Each contains two convolution layers with batch normalization \cite{ioffe2015batch} and two fully-connected layers. Strategy 1 reduces the feature size between the last convolution layer and the first fully-connected layer, while strategy 2 reduces the feature size between convolution layers. Each encoder uses a hyperbolic tangent (tanh) activation function after the last fully-connected layer to constrain the embedding values between -1 and 1. After testing convolutional kernels of size $k = \{3, 5, 7\}$, the value $k=3$ resulted in the lowest reconstruction errors. 

\subsubsection{Strategy 1}

A popular strategy for CNN architectures is to reduce the width and height but increase the depth of intermediate features as we go deeper into the network, see \cite{simonyan2014very, he2016deep}. The first strategy follows three heuristics:
\begin{enumerate}
	\item Apply half padding, such that the output dimension of intermediate convolution features remains the same.\label{enum:1}
	\item Apply max-pooling after each convolution step with a stride and kernel size of 2, reducing the feature size by a factor of 4. \label{enum:2}
	\item Double the square cuboid depth after each convolution step. \label{enum:3}
\end{enumerate}
The result of this strategy is that the size (the number of features in the intermediate representations) is reduced by two after every convolution operation. We present the square cuboid depth and dimension at all stages of the models in Table \ref{tab:encoder-large}. The feature size (row 3) is the product of square cuboid depth (number of channels) and the dimension of the intermediate features. 
\begin{table}[ht]
	\centering
	\begin{tabular}{ccccccc}
		\hline
		&     Input     &    Conv1    &    Conv2    & Unroll & FC1 & FC2 \\ \hline
		Square cuboid depth  &      512      &    1 024    &    2 048    & 32 768 & 128 & 16  \\
		Square cuboid width $\times$ height & $16\times 16$ & $8\times 8$ & $4\times 4$ &   1    &  1  &  1  \\
		Feature size      &    131 072    &   65 536    &   32 768    & 32 768 & 128 & 16  \\
		\% of parameters    &      NA       &     17      &     68      &   NA   & 15  &  0  \\ \hline
	\end{tabular}
	\caption{Large encoder model with 27 798 672 parameters}\label{tab:encoder-large}
\end{table}
In strategy 1, the convolution step accounts for most parameters. The steepest decrease in feature size occurs between the second convolution block and the first fully-connected layer (from 32 768 to 128). 

\subsubsection{Strategy 2}

For the second strategy, we follow a trial and error approach and attempt to restrict the number of parameters in the model. We retain heuristics \ref{enum:1} and \ref{enum:2} from strategy 1, but the depth of features decrease between each convolution block. 

\begin{table}[ht]
	\centering
	\begin{tabular}{ccccccc}
		\hline
		&     Input     &    Conv1    &    Conv2    & Unroll & FC1 & FC2 \\ \hline
		Square cuboid depth  &      512      &     48      &     16      &  256   & 16  &  8  \\
		Square cuboid width $\times$ height & $16\times 16$ & $8\times 8$ & $4\times 4$ &   1    &  1  &  1  \\
		Feature size      &    131 072    &    3 072    &     256     &  256   & 16  &  8  \\
		\% of parameters    &      NA       &     95      &      3      &   NA   &  2  &  0  \\ \hline
	\end{tabular}
	\caption{Small encoder model with 232 514 parameters}\label{tab:encoder-small}
\end{table}
In strategy 2, 95\% of the parameters are in the first convolution step. The feature size decreases steadily between each operation. 

\subsection{CRAE \& CBOW-CRAE decoders}

The output for the CRAE model is the reconstructed geographic data cuboid. The decoder in this model is the inverse operations of the encoder (deconvolutions and max-unpooling). The final activation function is sigmoid because the original inputs are between 0 and 1. We present the decoder operations for the large and the small decoders in Tables \ref{tab:decoder-large} and \ref{tab:decoder-small}. Recall that the input to the decoder is the embedding layer from the encoder.
\begin{table}[ht]
	\centering
	\begin{tabular}{ccccccc}
		\hline
		& Input & FC3 &  FC4   &    Roll     &   Deconv1   &    Deconv2    \\ \hline
		Square cuboid depth  &  16   & 128 & 32 768 &    2 048    &    1 024    &      512      \\
		Square cuboid width $\times$ height &   1   &  1  &   1    & $4\times 4$ & $8\times 8$ & $16\times 16$ \\
		Feature size      &  16   & 128 & 32 768 &   32 768    &   65 536    &    131 072    \\
		\% of parameters    &  NA   &  0  &   15   &     NA      &     68      &      17       \\ \hline
	\end{tabular}
	\caption{Large CRAE decoder model}\label{tab:decoder-large}
\end{table}
\begin{table}[ht]
	\centering
	\begin{tabular}{ccccccc}
		\hline
		& Input & FC3 & FC4 &    Roll     &   Deconv1   &    Deconv2    \\ \hline
		Square cuboid depth  &   8   & 16  & 256 &     16      &     48      &      512      \\
		Square cuboid width $\times$ height &   1   &  1  &  1  & $4\times 4$ & $8\times 8$ & $16\times 16$ \\
		Feature size      &  8     &  16   & 256 &     256     &    3 072    &    131 072    \\
		\% of parameters    &  NA   & 91  &  3  &     NA      &      2      &       0       \\ \hline
	\end{tabular}
	\caption{Small CRAE decoder model}\label{tab:decoder-small}
\end{table}

The CBOW-CRAE is a context to location model, so we select a fully-connected decoder, increasing from the embedding ($\gamma^*$) size $\ell$ to the geographic variable ($\gamma$) size $d$. In our experience, the decoder's exact dimensions did not significantly impact the reconstruction error, so we select the ascent dimensions (FC3 and FC4) to be the same as the fully-connected descent dimensions (FC1 and FC2). When there is no hidden layer from the embedding to the output (if there is only one fully-connected layer), the model is too linear to reconstruct the input data. When there is one hidden layer, the model is mainly able to reconstruct the data. Additional hidden layers did not significantly reduce the reconstruction error, so we select only one hidden layer in the decoder. Table \ref{tab:decoder-cbow-crae} presents the CBOW CRAE decoders in our implementation.
\begin{table}[ht]
	\centering
	\begin{tabular}{ccccc}
		\hline
		& Input & FC3 & FC4 &  \\ \hline
		Small model &   8   & 16  & 512 &  \\
		Large model &  16   & 128 & 512 &  \\ \hline
	\end{tabular}
	\caption{CBOW-CRAE decoders}\label{tab:decoder-cbow-crae}
\end{table}

\subsection{Comments on general hyperparameters and optimization strategy}\label{ss:comments}

We now offer a few comments on the training strategy. We split the postal codes into a training set and a validation set. Since the dataset is very large, we select a test set composed of only 5\% of the postal codes. We train the neural networks on a GeForce RTX 2080 8 GB GDDR6 graphics card and present the approximate training time later in this section. The batch size is the largest power of 2 that fits on the graphics card. We train the neural networks in PyTorch \cite{paszke2019pytorch} with the Adam optimization procedure from \cite{kingma2014adam}. We do not use weight decay (L2 regularization) since the model does not overfit. The initial learning rate for all models is 0.001 and decreases by a factor of 10 when losses stop decreasing for ten consecutive epochs. After five decreases, we stop the training procedure. 

The most significant issue during training is the saturation of initial hidden values (see, e.g., \cite{glorot2010understanding} for a discussion of the effect of neuron saturation and weight initialization). The encoder and the decoder's output are respectively tanh and sigmoid activations, which have horizontal asymptotes and small derivatives for high magnitude inputs. All models use batch normalization, without which the embeddings saturate quickly. Initializing the network with large weights, using the techniques from \cite{glorot2010understanding}, generated saturated embedding values of -1 or 1. To improve training, we initialize our models with very small weights, such that the average embedding value has a small variance and zero-centered. The neural network gradually increases the weights, so embeddings saturate less. 

The activation functions for intermediate layers are hyperparameters; we compare tanh and rectified linear unit (ReLU). The ReLU activation function generated the best performances, but the models did not converge for every initial seed. Our selected models use ReLU, but sometimes required restarting the training process with a different initial seed if the embeddings saturate. 

\subsection{Training results}

In this section, we provide results on the implementations of the four geographic embedding architectures, along with observations. Table \ref{tab:reconstruction-errors} presents the training and validation reconstruction errors, along with the training time, the number of parameters and the mean embedding values. 
\begin{table}[ht]
	\centering
	\begin{tabular}{cccccl}
		\hline
		& Training MSE & Validation MSE &  Time   & Parameters & Mean value \\ \hline
		Small CRAE    &  0.21299473  &   0.21207126   & 5 hours &  465 688   & 0.2051166     \\
		Large CRAE    &  0.21088413  &   0.20995240   & 3 days  & 55 622 416 & 0.6909323     \\
		Small CBOW-CRAE &  0.21833613  &   0.21715157   & 2 hours &  241 384   & 0.3463651     \\
		Large CBOW-CRAE &  0.21731463  &   0.21609975   & 2 days  & 27 866 896 & 0.1174563     \\ \hline
	\end{tabular}
	\caption{Reconstruction errors from architectures}\label{tab:reconstruction-errors}
\end{table}

One cannot directly compare the reconstruction errors for the classic CRAE and CBOW-CRAE since classic CRAE reconstructs $p^2$ as many values as CBOW-CRAE. The average reconstruction error for CRAE is smaller than for CBOW-CRAE, which could be because the output of CBOW-CRAE does not have a determined equivalent vector in the input data. {The CRAE model attempts to construct a one-to-one identity function for every neighbor because the input is identical to the output. On the other hand, CBOW-CRAE cannot exactly predict the values for a specific neighbor in the grid since there is no guarantee that the specific neighbor belongs to the same polygon as the central coordinate.} One also notices that the validation data's reconstruction error is smaller than the training data, which is atypical in machine learning. However, changing the initial seed for training and validation data changes this relationship, so one attributes this effect to the specific data split. This also means that the model does not overfit on the training data: if it did, then the training error would be much smaller than the validation error. The lack of overfit is a result of the bottleneck dimension being small (8 or 16 dimensions) with respect to the dimension of the input data (131 072).

For space and clarity reasons, we will not perform the explicit and implicit evaluations of embeddings. We do not find that one set of embeddings always performs better than the others, but find that the Large CBOW-CRAE behaves more appropriately, even if the reconstruction error is worse than for CRAE model. First, the average embeddings values for CBOW-CRAE models are closer to 0, which is desirable to increase the representation flexibility (especially within a GLM, because the range of embeddings is [-1, 1]). Attempts to manually correct these issues (normalization of embedding values after training) do not improve the quality of embeddings. In addition, the Large CBOW-CRAE had less saturated embedding dimensions, as we will discuss in the implicit evaluations. For these reasons, we continue our evaluation of embeddings with the Large CBOW-CRAE model, but we encourage researchers to experiment with other configurations. 

\subsection{Implicit evaluation of embeddings}\label{ss:implicit}

\begin{figure}[ht]
	\centering
	\begin{subfigure}[b]{0.32\textwidth}
		\centering
		\includegraphics[width = \textwidth]{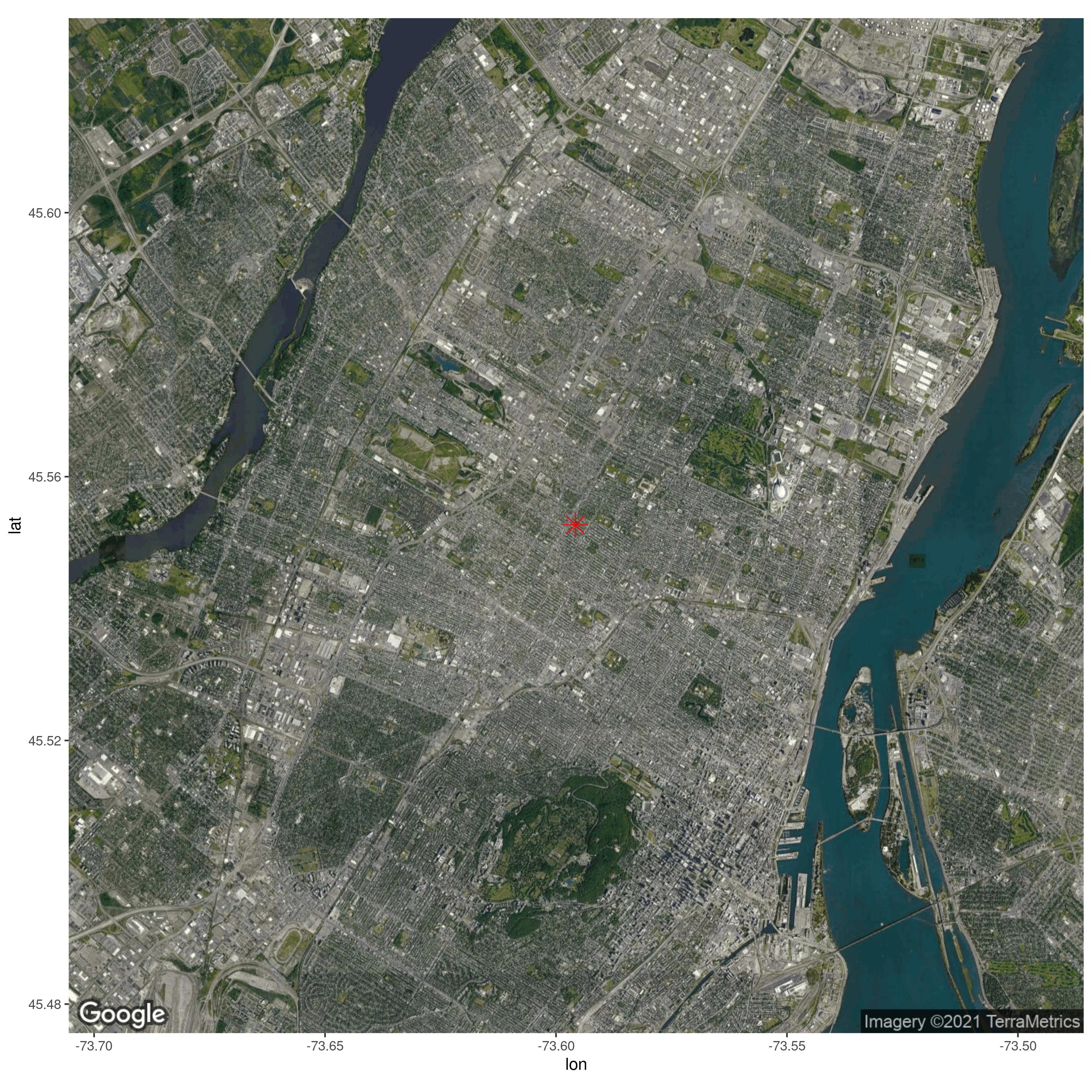}
		\caption{Empty map}
		\label{fig:embedding_0_zoom_12}
	\end{subfigure}
	\begin{subfigure}[b]{0.32\textwidth}
		\centering
		\includegraphics[width = \textwidth]{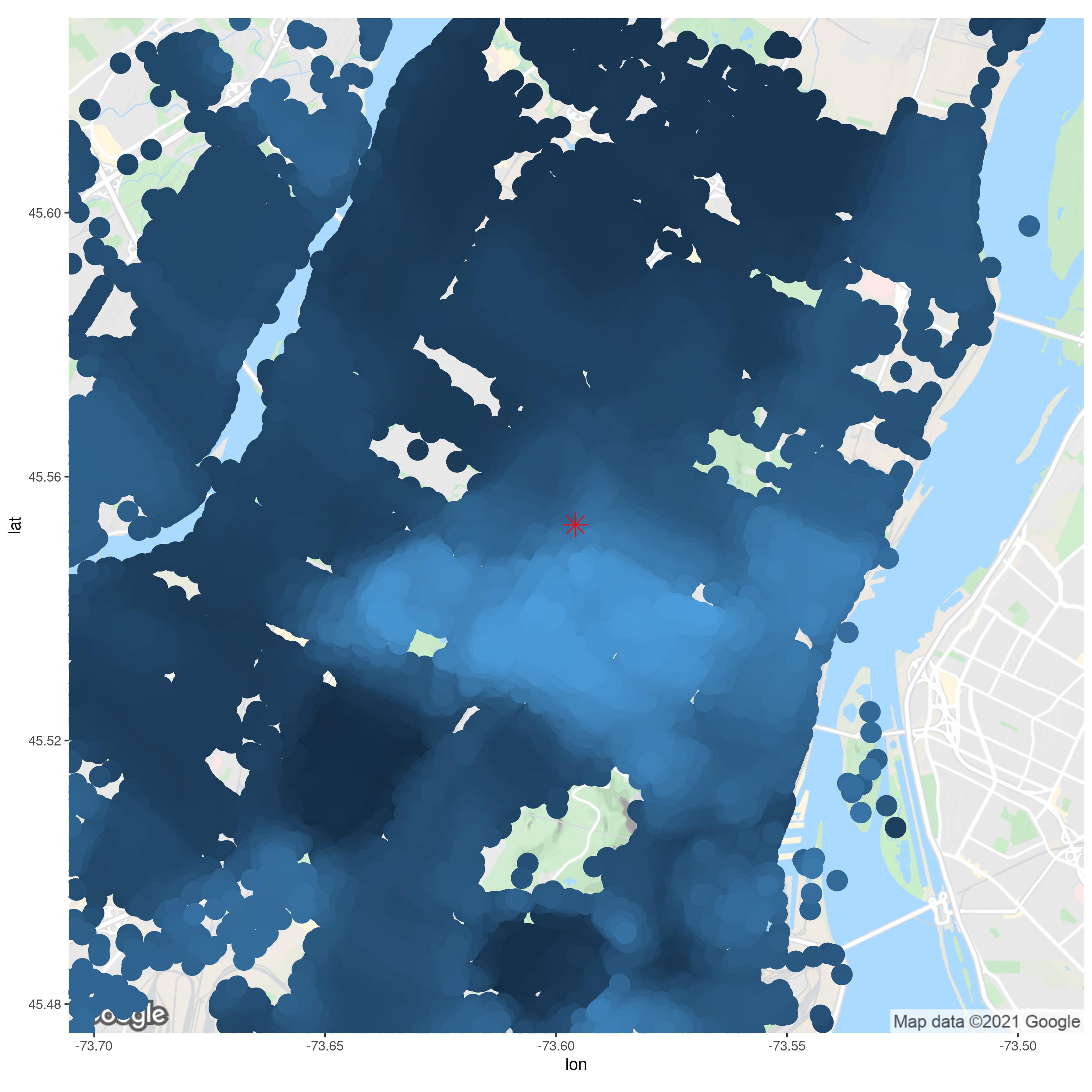}
		\caption{Dimension 1}
		\label{fig:embedding_1_zoom_12}
	\end{subfigure}
	\begin{subfigure}[b]{0.32\textwidth}
		\centering
		\includegraphics[width = \textwidth]{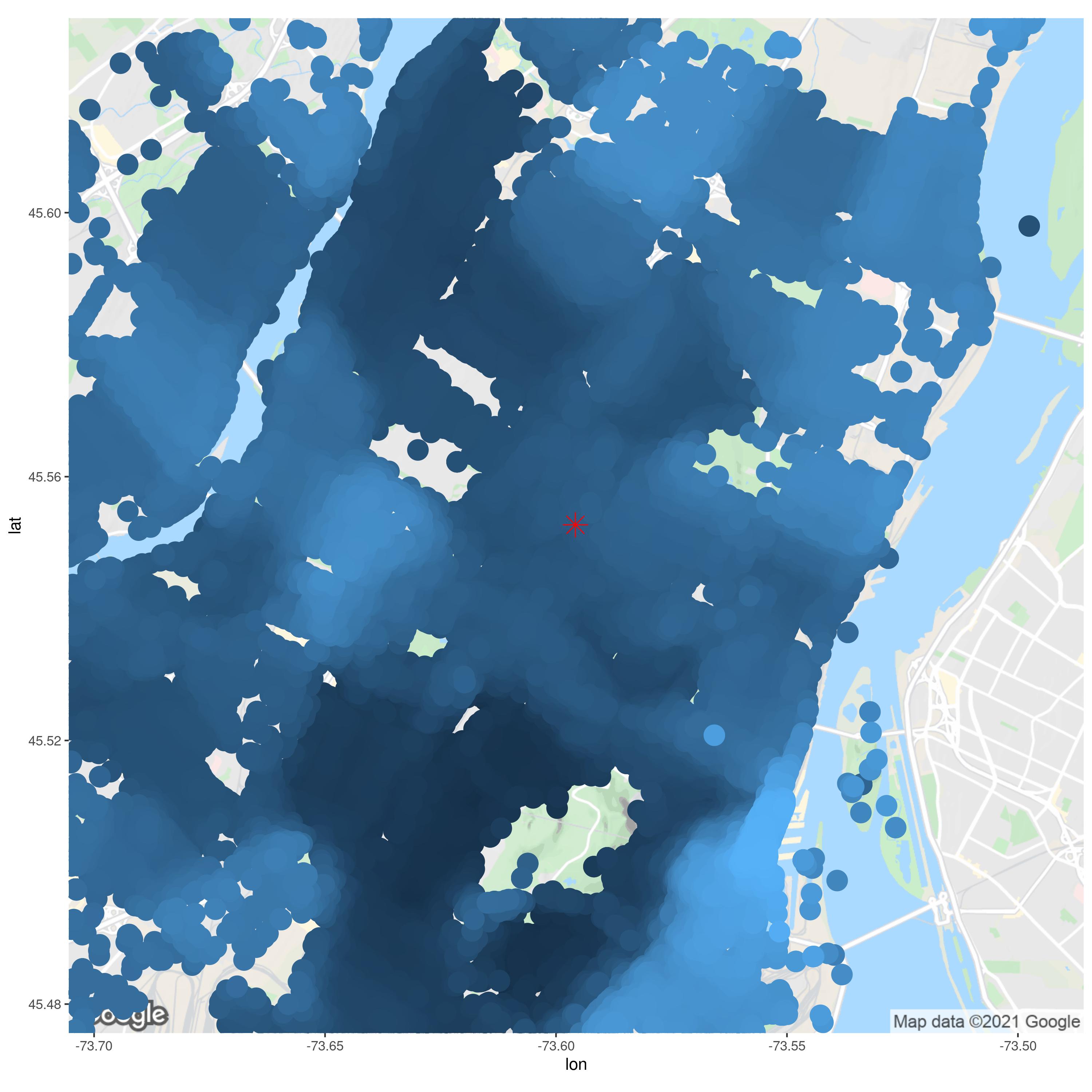}
		\caption{Dimension 2}
		\label{fig:embedding_2_zoom_12}
	\end{subfigure}
	\caption{Visually inspecting embedding dimensions on a map for the Island of Montréal}
	\label{fig:embedding_zoom_12}
\end{figure}

We now implicitly evaluate if the 16 dimensions of the embeddings (generated by the Large CBOW-CRAE model) follows attribute \ref{att:tfl1} (TFL). Figure \ref{fig:embedding_zoom_12} presents an empty map for a location in Montréal (to identify points of interest), along with two embedding dimensions. The red star is the same coordinate as Figure \ref{fig:span}. The map includes two rivers (in blue), an airport (bottom left), a mountain (bottom middle), and other parks. These points of interest typically have few surrounding postal codes, so the maps of embedding are less dense than heavily populated areas. The maps of embeddings include no legend because the magnitude of embeddings is irrelevant (regression weights will account for their scale). 
Not only are the embeddings smooth, but different dimensions learn different patterns. Recall that a polygon-based embedding model will learn the same shape (subject to the shape of polygons). Since models based on the geographic data cuboid depend on coordinates, the embeddings' shapes are more flexible. Inspecting Figures \ref{fig:embedding_1_zoom_12} and \ref{fig:embedding_2_zoom_12} around the red star, we observe that the embedding values form different shapes, and these shapes are different from the FSA polygons of Figure \ref{fig:grid}, validating attribute 2 (coordinate). 

\begin{figure}[ht]
	\centering
	\begin{subfigure}[b]{0.24\textwidth}
		\centering
		\includegraphics[width = \textwidth]{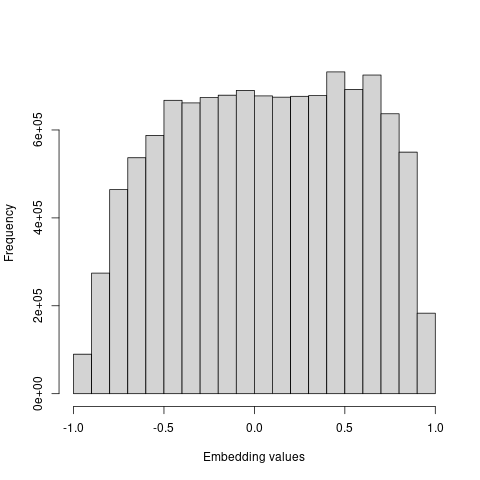}
		\caption{Other dimensions}
		\label{fig:hist-all}
	\end{subfigure}
	\begin{subfigure}[b]{0.24\textwidth}
		\centering
		\includegraphics[width = \textwidth]{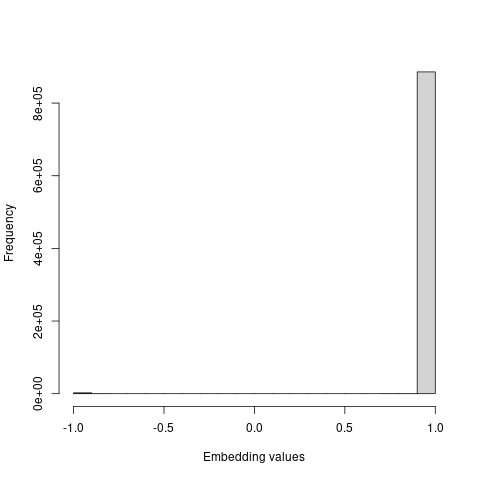}
		\caption{Dimension 3}
		\label{fig:hist-3}
	\end{subfigure}
	\begin{subfigure}[b]{0.24\textwidth}
		\centering
		\includegraphics[width = \textwidth]{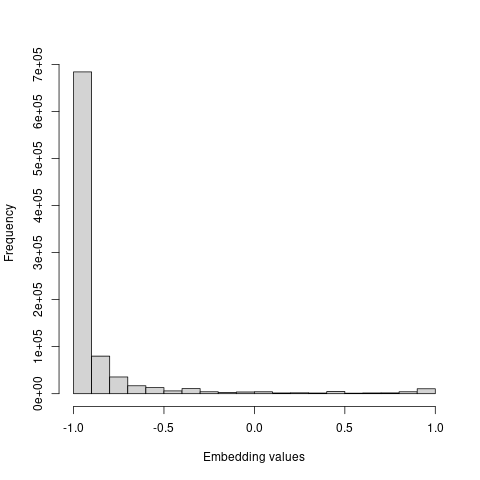}
		\caption{Dimension 4}
		\label{fig:hist-4}
	\end{subfigure}
	\begin{subfigure}[b]{0.24\textwidth}
		\centering
		\includegraphics[width = \textwidth]{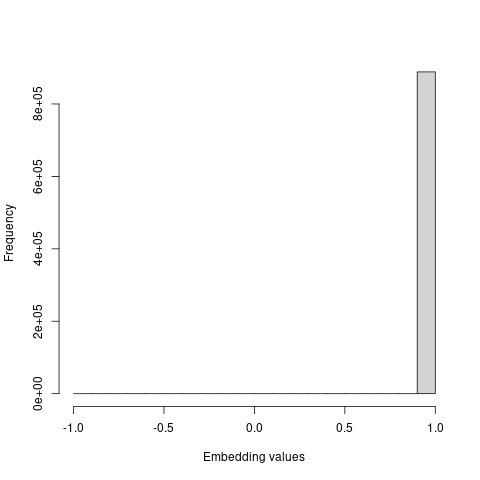}
		\caption{Dimension 6}
		\label{fig:hist-6}
	\end{subfigure}
	\caption{Histogram of embedding dimensions. Dimensions 3 and 6 are saturated}
	\label{fig:hist-embeddings}
\end{figure}

Visually inspecting the embeddings diagnosed an issue of embedding saturation, as discussed in Section \ref{ss:comments}. Saturated embeddings all equal the value -1 or 1, and because of the flat shape of the hyperbolic tangent activation function, the gradients of the model weights are too small for the optimizer to correct the issue. In Figure \ref{fig:hist-embeddings}, we present the histogram of the embedding values for dimensions 3, 4, 6, and the remaining dimensions combined. A good embedding will have values distributed along the entire support [-1, 1], as presented in Figure \ref{fig:hist-all}. Dimensions 3 and 6 are heavily saturated, having most values equal exactly 1 or -1. If one includes embedding dimension 6 in a GLM, the feature will replicate the model intercept exactly (and cause collinearity issues if we include both). Embedding dimension 3 is less problematic since there is a small proportion of -1 values but still exhibits saturation. Most embedding values from dimension 4 are saturated at -1 but also have other values on the support of the activation function, which could provide useful information. We decide to discard embedding dimensions 3 and 6 for our applications. 

We visually conclude that the remaining embedding dimensions satisfy all desirable attributes \ref{att:tfl1} to \ref{att:external3} of geographic embeddings.
    % !TeX spellcheck = en_US
\section{Applications}\label{sec:application}

In this section, we present applications of predictive models for accident frequency. For all applications, we organize datasets as in Table \ref{tab:spatial-db-embedding}. All applications are on Canadian data, so we project the coordinates to Atlas Lambert. Each observation also has geographic embeddings to describe the geography of the observation. The geographic embeddings remain the same, no matter the application: they correspond to the optimal Large CBOW-CRAE model trained in Section \ref{sec:implementation}. Finally, we have typical non-geographic information like traditional actuarial features, the exposure and the response variable. 

\begin{table}[ht]
	\centering
	\begin{tabular}{@{}rllllllllll@{}}
		\toprule
		Index & \multicolumn{2}{c}{Coordinates} &         \multicolumn{3}{c}{Geographic embeddings}          & \multicolumn{3}{c}{Other features} & Exposure   & Response \\ \cmidrule(r){1-1}\cmidrule(r){2-3}\cmidrule(r){4-6}\cmidrule(r){7-9}\cmidrule(r){10-10}\cmidrule(r){11-11}
		$i$ & $lon_{i}$ & $lat_{i}$           & $\gamma^*_{i,1}$ & \dots    & $\gamma^*_{i,\ell}$ & $x_{i1}$ & \dots    & $x_{ip}$ & $\omega_i$ & $y_i$    \\ \midrule
		1 & $lon_{1}$ & $lat_{1}$           & $\gamma^*_{1,1}$ & \dots    & $\gamma^*_{1,\ell}$ & $x_{11}$ & \dots    & $x_{1p}$ & $\omega_1$ & $y_1$    \\
		2 & $lon_{2}$ & $lat_{2}$           & $\gamma^*_{2,1}$ & \dots    & $\gamma^*_{2,\ell}$ & $x_{21}$ & \dots    & $x_{2p}$ & $\omega_2$ & $y_2$    \\
		$\vdots$ & $\vdots$  & $\vdots$            & $\vdots$        & $\ddots$ & $\vdots$           & $\vdots$ & $\ddots$ & $\vdots$ & $\vdots$   & $\vdots$ \\
		$n$ & $lon_{n}$ & $lat_{n}$           & $\gamma^*_{n,1}$ & \dots    & $\gamma^*_{n,\ell}$ & $x_{n1}$ & \dots    & $x_{np}$ & $\omega_n$ & $y_n$    \\ \bottomrule
	\end{tabular}
	\caption{Example of a geographic dataset with geographic embeddings}\label{tab:spatial-db-embedding}
\end{table}

\subsection{Dataset 1: Car accident counts}\label{ss:car}

The first application is to predict car accident frequency in different postal codes on the island of Montréal between 2012 and 2017. The data comes from the \emph{Société de l'assurance automobile du Québec} (SAAQ), the public insurance company for the province of Québec. The city of Montréal publishes a modified dataset including the coordinates of the closest intersection of the car accident (\url{https://donnees.montreal.ca/ville-de-montreal/collisions-routieres}).
We then allocate the accidents to the nearest postal code. We organize data as in Table \ref{tab:spatial-db-embedding}, with each observation representing one postal code. We use the years 2012 to 2016 for the training dataset, which contains 116 118 car accidents. We keep the year 2017 as an out-of-sample test dataset, containing 19 997 car accidents. 

The model for the first application is a generalized additive model with a Poisson response to model the accident count. The relationship between the expected value of the response variable $Y$, the coordinates and the geographic embeddings is given by
\begin{equation}\label{eq:car-gam}
	\ln(E[Y_i]) = \beta_0 + \underbrace{\vphantom{\sum_{k = 1}^{\ell}}f_k(lon_i, ~lat_i)}_{\text{spline component}} + \underbrace{\sum_{j = 1}^{14}\gamma^*_{ij} \beta_{j},}_{\text{embedding component}} ~ i = 1, \dots, n,
\end{equation}
where $f_k$ is a bivariate spline function with $k\times k$ knots (in our application, a bivariate tensor product smooth, see \cite{wood2012mgcv} for details). The number of knots in the splines is a hyperparameter controlling flexibility. Model \eqref{eq:car-gam} contains a smooth geographic component from the bivariate spline, along with a linear geographic embedding component. One notices an advantage of geographic embeddings: instead of dealing with geographic coordinates directly, they capture geographic effects linearly with regression coefficients. Since geographic embeddings satisfy TFL, a linear combination of geographic embeddings also follows TFL. Geographic embeddings do not entirely replace geographic models: it is simple to combine traditional geographic models with the embeddings. 

We compare the models with only the embedding component ($k = 0$ with embeddings), models with only spline components ($k > 0$ without embeddings), and combinations of embeddings and splines ($k > 0$ with embeddings). Table \ref{tab:results-car} presents the training deviance, test deviance, model degrees of freedom (DoF) and training time in seconds using the restricted maximum likelihood method using the mgcv R package. The DoF from the embedding component is constant at 14. For the spline component, we provide the effective DoF, which corresponds to the trace of the hat matrix of spline predictors; see \cite{wood2012mgcv} for details. The training times are based on a computer with Intel{\textregistered}~Core{\texttrademark}~i5-7600K CPU @ 3.80GHz.
\begin{table}[ht]
	\centering
	\begin{tabular}{@{}ccccccccc@{}}
		\toprule
		& \multicolumn{4}{c}{Without embeddings} & \multicolumn{4}{c}{With embeddings $\boldsymbol{\gamma}^*$} \\
		\cmidrule(r){1-1}\cmidrule(r){2-5}\cmidrule(r){6-9}
		$k$ & Training & Test  &   DoF   &  Time (s)  & Training & Test  &   DoF   &            Time (s)             \\
		\cmidrule(r){1-1}\cmidrule(r){2-5}\cmidrule(r){6-9}
		0  &    --    &  --   &   --   &     --     &  153592  & 72419 & 15.00  &                0                \\
		3  &  156424  & 73247 &  8.98  &     2      &  153305  & 72349 & 20.53  &                5                \\
		5  &  156106  & 73237 & 18.39  &     3      &  152921  & 72205 & 32.58  &                5                \\
		8  &  153701  & 72353 & 45.25  &     10     &  151369  & 71550 & 59.19  &               11                \\
		10  &  152050  & 71822 & 78.03  &     74     &  150856  & 71374 & 80.13  &               13                \\
		15  &  149770  & 70859 & 155.39 &     66     &  149099  & 70612 & 166.49 &               71                \\
		20  &  147548  & 69818 & 263.76 &    341     &  146921  & 69589 & 275.58 &               148               \\
		25  &  144789  & 68602 & 410.49 &    328     &  144379  & \textbf{68476} & 420.32 &               607               \\ \bottomrule
	\end{tabular}
	\caption{Performance comparison for car accident count models}\label{tab:results-car}
\end{table}

One observes that for fixed $k$, including embeddings in the model decreases both train and test deviance. Increasing $k$, the number of knots in the spline component increases the effective DoF more than linearly. On the other hand, including the embedding component increases the effective DoF by 14 since embeddings' dimension remains the same. The number of knots needed in a spline model without embeddings to outperform the GLM without splines is $k = 7$, with 36.81 effective DoF. To improve model performance, it is generally more advantageous to add the embedding component than to increase the number of knots in the splines (based on the increase in effective DoF and the increase in training time). 

To evaluate the embeddings extrinsically, one can study the statistical significance of the regression parameters. Table \ref{tab:car-p-values} presents the \emph{p-values} for the regression parameters in the car accident count GLM model ($k = 0$ with embeddings). One notices that 13 embeddings have statistically significant regression at the 0.05 threshold, with 11 also statistically significant at the $10\times 10^{-5}$ level, which means that geographic embeddings capture useful features. 
\begin{table}[ht]
	\centering
	\begin{tabular}{@{}clclclcl@{}}
		\toprule
		Parameter & \emph{p-value} &  Parameter  & \emph{p-value} &  Parameter   & \emph{p-value} &  Parameter   & \emph{p-value} \\
		\cmidrule(r){1-2}\cmidrule(r){3-4}\cmidrule(r){5-6}\cmidrule(r){7-8}
		$\beta_0$ & 0.00000        & $\beta_{4}$ & 0.00000        & $\beta_{8}$  & 0.00355        & $\beta_{12}$ & 0.00000        \\
		$\beta_1$ & 0.00000        & $\beta_{5}$ & 0.00000        & $\beta_{9}$  & 0.00000        & $\beta_{13}$ & 0.00000        \\
		$\beta_2$ & 0.00000        & $\beta_{6}$ & 0.00752        & $\beta_{10}$ & 0.00000        & $\beta_{14}$ & 0.00000        \\
		$\beta_3$ & 0.00000        & $\beta_{7}$ & 0.00000        & $\beta_{11}$ & 0.07549        &              &                \\ \bottomrule
	\end{tabular}
	\caption{\textit{p}-values for car count GLM model}\label{tab:car-p-values}
\end{table}

%Coefficients:
%Estimate Std. Error z value Pr(>|z|)    
%(Intercept) -2.42978    0.07904 -30.741  < 2e-16 ***
%V1          -0.14465    0.01707  -8.476  < 2e-16 ***
%V2          -0.32370    0.03108 -10.414  < 2e-16 ***
%V4          -0.80181    0.07212 -11.118  < 2e-16 ***
%V5          -0.70094    0.01672 -41.916  < 2e-16 ***
%V7           0.19644    0.01744  11.266  < 2e-16 ***
%V8          -0.03357    0.01256  -2.673  0.00752 ** 
%V9          -0.41733    0.02493 -16.737  < 2e-16 ***
%V10          0.05339    0.01831   2.915  0.00355 ** 
%V11         -0.17249    0.01910  -9.031  < 2e-16 ***
%V12          0.19129    0.01506  12.705  < 2e-16 ***
%V13          0.03287    0.01849   1.777  0.07549 .  
%V14         -0.50265    0.02001 -25.117  < 2e-16 ***
%V15          0.27453    0.02665  10.301  < 2e-16 ***
%V16         -0.35331    0.03384 -10.442  < 2e-16 ***

\subsection{Dataset 2: Fire incident counts}\label{ss:fire}

The second application predicts fire incidents in different postal codes in Toronto between 2011 and July 2019. The data comes from City of Toronto Open Data (\url{https://open.toronto.ca/dataset/fire-incidents/}).
The dataset contains the coordinates of the nearest major intersection of fire incidents, which we allocate to the nearest postal code. We organize data as in Table \ref{tab:spatial-db-embedding}, with each observation representing one postal code. We use the years 2011 to 2017 for the training dataset, containing 12 540 incidents. We keep the year 2018 and the first half of 2019 as out-of-sample test dataset, containing 2 918 incidents. 

The model for the second application is also a generalized additive model with a Poisson response. The relationship between the expected value of the response variable $Y$, the coordinate predictors and the geographic embeddings is still \eqref{eq:car-gam}. Table \ref{tab:results-fire} presents the training and test deviance, along with effective DoF and training time (still based on a computer with Intel{\textregistered}~Core{\texttrademark}~i5-7600K CPU @ 3.80GHz.). One draws similar conclusions to the first application: increasing $k$ reduces the train and test deviance, and so does including geographic embeddings in the model. The number of knots needed in a spline model without embeddings to outperform the GLM without splines is $k = 8$, with 49.8 effective DoF. 
\begin{table}[ht]
	\centering
	\begin{tabular}{@{}ccccccccc@{}}
		\toprule
		& \multicolumn{4}{c}{Without embeddings} & \multicolumn{4}{c}{With embeddings $\boldsymbol{\gamma}^*$} \\
		\cmidrule(r){1-1}\cmidrule(r){2-5}\cmidrule(r){6-9}
		$k$ & Training & Test  &   DoF   &  Time (s)  & Training & Test  &   DoF   &            Time (s)             \\ \cmidrule(r){1-1}\cmidrule(r){2-5}\cmidrule(r){6-9}
		0  &    --    &  --   &   --   &     --     &  36323   & 13979 &   15   &                0                \\
		3  &  37622   & 14360 &  6.94  &     2      &  36161   & 13940 & 21.57  &                5                \\
		5  &  37084   & 14186 & 19.45  &     5      &  36019   & 13897 & 30.41  &               12                \\
		8  &  36480   & 13985 & 49.80  &     28     &  35840   & 13846 & 52.02  &               20                \\
		10  &  36371   & 13937 & 67.23  &     64     &  35628   & 13777 & 76.89  &               47                \\
		15  &  35414   & 13758 & 147.71 &    176     &  34814   & 13612 & 157.89 &               206               \\
		20  &  34461   & 13500 & 251.97 &    446     &  34046   & 13420 & 256.66 &               410               \\
		25  &  33713   & 13401 & 364.21 &    861     &  33312   & \textbf{13319} & 369.28 &              1413               \\ \bottomrule
	\end{tabular}
	\caption{Performance comparison for fire incident models}\label{tab:results-fire}
\end{table}

In Table \ref{tab:fire-p-values}, we provide the \emph{p-values} for the fire incident count GLM model ($k = 0$ with embeddings). One notices that ten embeddings have statistically significant regression at the 0.05 threshold, with seven also statistically significant at the $10\times 10^{-5}$ level. In the car accident application, the only regression coefficient that was not significant at the 0.05 level was $\beta_{11}$, which is highly significant for the fire incidents frequency model. %Since the embeddings compress socio-demographic information; one expects that not all embeddings are significant for different applications. 

\begin{table}[ht]
	\centering
	\begin{tabular}{@{}clclclcl@{}}
		\toprule
		Parameter & p-value &  Parameter  & p-value &  Parameter   & p-value &  Parameter   & p-value \\ \cmidrule(r){1-2}\cmidrule(r){3-4}\cmidrule(r){5-6}\cmidrule(r){7-8}
		$\beta_0$ & 0.00000 & $\beta_{4}$ & 0.00000 & $\beta_{8}$  & 0.13197 & $\beta_{12}$ & 0.11907 \\
		$\beta_1$ & 0.00000 & $\beta_{5}$ & 0.00000 & $\beta_{9}$  & 0.00808 & $\beta_{13}$ & 0.00000 \\
		$\beta_2$ & 0.05736 & $\beta_{6}$ & 0.10260 & $\beta_{10}$ & 0.01111 & $\beta_{14}$ & 0.00000 \\
		$\beta_3$ & 0.00000 & $\beta_{7}$ & 0.00000 & $\beta_{11}$ & 0.00203 &              &         \\ \bottomrule
	\end{tabular}
	\caption{\textit{p}-values for fire count model}\label{tab:fire-p-values}
\end{table}

The applications of datasets 1 and 2 compare the performance of geographic splines and geographic embeddings. One observes that GAMs without embeddings outperform GLMs with embeddings once the number of knots was sufficiently high. This finding is not surprising: the geographic embeddings in Section \ref{sec:implementation} depend on socio-demographic variables. Other geographic effects (meteorological, ecological, topological) could also affect the geographic risk. However, adding the embeddings to the model increased the performance, so that models should include embeddings for both datasets. One concludes that the geographic distribution of the population partially explains the geographic distribution of car accidents and fire incidents. Since residual effects (captured by the GAM splines) improve the model, socio-demographic information does not entirely explain the geographic distribution of risk. These conclusions hold for two highly populated cities (Montréal and Toronto) and for two predictive tasks that are not directly related to insurance, but which model P\&C perils (car accident and fire incident counts). 

\subsection{Dataset 3: Home insurance}\label{ss:home}

The third dataset contains historical losses of home insurance contracts for a portfolio in the province of Québec, Canada, between 2007 and 2011. The data comes from a large Canadian P\&C insurance company and contains over 2 million observations. The home insurance contract covers six perils, including \texttt{fire}, \texttt{water} damage, sewer backup (\texttt{SBU}), wind \& hail (\texttt{W\&H}), \texttt{theft} and a final category called \texttt{other}. The dataset provides the house's postal code for each observation, so we set the coordinates as the central point of the postal code and extract the embeddings from that same postal code. We also have traditional actuarial variables describing the house and the customer for each contract, along with the contract's length (exposure $\omega$, treated as offsets in our models). We organize data as in Table \ref{tab:spatial-db-embedding}, with each observation representing one insurance contract for one or fewer years. For illustration purposes, we select four traditional variables in the models, including $x_1$: age of the building, $x_2$: age of the client, $x_3$: age of the roof, and $x_4$: building amount. 

The third application uses a GAM with a Poisson response to model the home claim frequency. The relationship between the response variable $Y$ and the traditional variables, the geographic embeddings and the exposure is
\begin{equation}\label{eq:home-gam}
	\ln(E[Y_i]) = \beta_0 + \ln \omega + \underbrace{\sum_{j = 1}^{4}x_{ij} \alpha_{j}}_{\text{traditional component}} + \underbrace{\vphantom{\sum_{k = 1}^{\ell}}f_k(lon_i, ~lat_i)}_{\text{spline component}} + \underbrace{\sum_{j = 1}^{14}\gamma^*_{ij} \beta_{j},}_{\text{embedding component}} ~ i = 1, \dots, n.
\end{equation}
The training times provided for all home insurance dataset comparisons are based on a computer with two Intel\textregistered~Xeon\textregistered~Processor E5-2683 v4 @2.10GHz (about 3.7 times faster than the i5 processor when running at full capacity). 

%For this reason, the conclusions from the models in Section \ref{ss:car} and \ref{ss:fire} may change. \todo{Pas clair} 

%Intact : under 2\%, under 1\% for every peril
%Fire: 0.03154339 per postal code
%Car: 0.2818771 per postal code

\subsubsection{Home accident frequency in Montréal}\label{sss:home-mtl}

In the first comparison with the home insurance dataset, we attempt to predict the total claim frequency for an insurance contract on the island of Montréal. The training data are calendar year losses for 2007 to 2010, while the test data are the calendar year losses for 2011. Table \ref{tab:results-home-mtl} presents the training and test deviance, along with effective DoF and training time. 
\begin{table}[ht]
	\centering
	\begin{tabular}{@{}ccccccccc@{}}
		\toprule
		& \multicolumn{4}{c}{Without embeddings} & \multicolumn{4}{c}{With embeddings $\boldsymbol{\gamma}^*$} \\
		\cmidrule(r){1-1}\cmidrule(r){2-5}\cmidrule(r){6-9}
		$k$ & Training & Test  &  DoF   &  Time (s)   & Training & Test  &  DoF   &             Time (s)             \\
		\cmidrule(r){1-1}\cmidrule(r){2-5}\cmidrule(r){6-9}
		0  &    --    &  --   &  --   &     --      &  66013   & \textbf{14383} &  19   &                58                \\
		3  &  66149   & 14390 & 6.69  &     242     &  65952   & 14399 & 23.70 &               483                \\
		5  &  65991   & 14400 & 16.49 &     108     &  65886   & 14402 & 33.61 &               1553               \\
		8  &  65838   & 14390 & 34.00 &    1306     &  65778   & 14388 & 48.65 &               1024               \\
		10  &  65766   & 14389 & 46.03 &    2201     &  65733   & 14388 & 58.21 &               1540               \\
		15  &  65691   & 14389 & 64.44 &    2533     &  65683   & 14391 & 73.42 &               3617               \\
		20  &  65652   & 14386 & 75.37 &    7733     &  65651   & 14387 & 82.29 &              12368               \\
		25  &  65644   & 14386 & 80.36 &   50902     &  65642   & 14387 & 86.39 &              50763               \\ \bottomrule
	\end{tabular}
	\caption{Performance comparison for home total claim frequency models (Montréal)}\label{tab:results-home-mtl}
\end{table}

The best model is the GLM with embeddings. Including splines with embeddings does not improve the model: although the training deviance decreases, the test deviance increases. One concludes that when smoothing the frequency of rare events with limited observations, splines learn patterns that overfit on observed loss experience and do not learn geographic effects that generalize to new observations. These conclusions are in contrast to the applications of Sections \ref{ss:car} and \ref{ss:fire} (which had higher frequencies), where increasing the knots decreased the test deviance. 

\subsubsection{Home claim frequency in the entire portfolio}\label{sss:home-qc}

In the second comparison with the home insurance dataset, we train the ratemaking models on the entire province of Québec. This application is most representative of the ratemaking models in practice. We note that dataset 1 (Montréal) covered 365 km\textsuperscript{2}, dataset 2 (Toronto) covered 630 km\textsuperscript{2}, while this application covers almost 1 400 000 km\textsuperscript{2}, so the spline component will require a much larger number of knots to replicate the flexibility of the models in Sections \ref{ss:car}, \ref{ss:fire} and \ref{sss:home-mtl}. Also, most of the area in Québec is uninhabited, so much of the flexibility is wasted on locations with no observations. The geographic embeddings in this application are the same as the previous applications, so the models with geographic embeddings have the same flexibility without significantly increasing the effective DoF. Table \ref{tab:results-home-qc} presents the training and test deviance, along with effective DoF and training time for models. Note that we omit $k = 25$ since training the model would require too much RAM on the compute server. 
\begin{table}[ht]
	\centering
	\begin{tabular}{@{}ccccccccc@{}}
		\toprule
		& \multicolumn{4}{c}{Without embeddings} & \multicolumn{4}{c}{With embeddings $\boldsymbol{\gamma}^*$} \\
		\cmidrule(r){1-1}\cmidrule(r){2-5}\cmidrule(r){6-9}
		$k$ & Training & Test  &  DoF    &  Time (s)   & Training & Test  &  DoF    &             Time (s)             \\
		\cmidrule(r){1-1}\cmidrule(r){2-5}\cmidrule(r){6-9}
		0  &    --    &  --   &  --    &     --      &  332577  & \textbf{80663} &  19    &               312                \\
		3  &  333231  & 80766 & 7.18   &    4604     &  332422  & 80785 & 26.16  &              18158               \\
		5  &  332779  & 80767 & 18.63  &    5709     &  332251  & 80780 & 35.18  &              9809                \\
		8  &  332346  & 80837 & 43.97  &   17775     &  331948  & 80777 & 60.43  &              28823               \\
		10  &  331791  & 80771 & 64.89  &   41212     &  331550  & 80748 & 82.86  &              52041               \\
		15  &  331174  & 80853 & 126.60 &   37885     &  331025  & 80825 & 138.64 &              56316               \\
		20  &  330858  & 80852 & 181.65 &   59935     &  330763  & 80825 & 191.97 &              51102               \\\bottomrule
	\end{tabular}
	\caption{Performance comparison for home total claim frequency models (Province of Québec)} \label{tab:results-home-qc}
\end{table}

Once again, the best model is the GLM with embeddings. One observes a decreasing trend in the training deviance as $k$ increases, but an increase in test deviance follows. One concludes that geographic embeddings capture all significant geographic risk, and splines overfit the training data without generalizing to new observations. 

\subsubsection{Extrinsic evaluation for home claim frequency}\label{sss:home-qc-perils}

We now extrinsically evaluate geographic embeddings for predicting home insurance claim frequency. The home insurance contracts in our dataset cover six perils. We train seven GLM models with geographic embeddings and the four traditional variables: one for the total claims frequency and six for the claim frequency decomposed by individual peril. In Table \ref{tab:p-values-home}, we present the \emph{p-values} for the regression coefficients in the seven models, along with the number of times that the regression coefficients were significant at the $0.05$ level (column \#). 
\begin{table}[ht]
	\centering
	\begin{tabular}{@{}lllllllll@{}}
		\hline
		& Total  & \texttt{Fire} & \texttt{Theft} & \texttt{W\&H} & \texttt{Water} & \texttt{SBU} & \texttt{Other} & \#  \\ \hline
		$\beta_{0}$  & 0.0000 & 0.1575        & 0.0009         & 0.0235        & 0.0000         & 0.0009       & 0.0000         & 6  \\
		$\beta_{1}$  & 0.0172 & 0.0204        & 0.0000         & 0.5152        & 0.0200         & 0.0001       & 0.0003         & 6  \\
		$\beta_{2}$  & 0.0000 & 0.8525        & 0.0198         & 0.0000        & 0.0000         & 0.0000       & 0.0000         & 6  \\
		$\beta_{3}$  & 0.0002 & 0.6756        & 0.0009         & 0.6661        & 0.3390         & 0.2551       & 0.0004         & 3  \\
		$\beta_{4}$  & 0.0169 & 0.0006        & 0.0000         & 0.0000        & 0.0000         & 0.0052       & 0.0027         & 7  \\
		$\beta_{5}$  & 0.0000 & 0.0050        & 0.0000         & 0.0000        & 0.0000         & 0.0000       & 0.0000         & 7  \\
		$\beta_{6}$  & 0.0033 & 0.0595        & 0.0018         & 0.0000        & 0.5091         & 0.0227       & 0.1292         & 4  \\
		$\beta_{7}$  & 0.0000 & 0.5498        & 0.0000         & 0.0000        & 0.3012         & 0.1538       & 0.0000         & 4  \\
		$\beta_{8}$  & 0.1695 & 0.0155        & 0.2429         & 0.0000        & 0.0007         & 0.4392       & 0.0040         & 4  \\
		$\beta_{9}$  & 0.0063 & 0.7287        & 0.6080         & 0.0033        & 0.0662         & 0.1104       & 0.8288         & 2  \\
		$\beta_{10}$ & 0.0000 & 0.3631        & 0.0001         & 0.2265        & 0.0000         & 0.5243       & 0.0009         & 4  \\
		$\beta_{11}$ & 0.0690 & 0.0032        & 0.0000         & 0.7296        & 0.0000         & 0.0051       & 0.9000         & 4  \\
		$\beta_{12}$ & 0.0000 & 0.0038        & 0.2538         & 0.2029        & 0.0000         & 0.0002       & 0.0000         & 5  \\
		$\beta_{13}$ & 0.0000 & 0.0000        & 0.0000         & 0.8895        & 0.0000         & 0.0012       & 0.0000         & 6  \\
		$\beta_{14}$ & 0.5908 & 0.0172        & 0.0000         & 0.0000        & 0.7580         & 0.0000       & 0.0318         & 5  \\
		$\alpha_{1}$ & 0.0028 & 0.0001        & 0.0514         & 0.0000        & 0.2503         & 0.0000       & 0.0000         & 5  \\
		$\alpha_{2}$ & 0.0000 & 0.0000        & 0.0000         & 0.5023        & 0.0000         & 0.0000       & 0.0000         & 6  \\
		$\alpha_{3}$ & 0.0000 & 0.2040        & 0.2510         & 0.0000        & 0.0005         & 0.0000       & 0.0004         & 5  \\
		$\alpha_{4}$ & 0.0000 & 0.1022        & 0.0000         & 0.5681        & 0.0000         & 0.0000       & 0.0000         & 5  \\ \hline
	\end{tabular}
	\caption{\emph{p-values} for the GLMs by peril}\label{tab:p-values-home}
\end{table}

For clarity and simplicity, we use the abuse of terminology \emph{significant embeddings} to mean that the regression coefficient associated with an embedding dimension is statistically significant, based on the \emph{p-value}. One notices that the 12 embedding dimensions are significant for the majority of models, and that regression coefficients $\beta_4$ and $\beta_5$ are significant for all models. The signs of the regression coefficients (not shown for space) are not the same for every peril, meaning some embedding dimensions increase the expected claim frequency for certain perils but decrease it for other perils. 

The peril GLMs with the most significant geographic embeddings are \texttt{theft} and \texttt{other}. Since the embeddings capture socio-demographic effects, it makes sense that the \texttt{theft} peril, which one mostly associates with socio-demographic actions, has the highest number of significant embeddings. The authors have no knowledge of the contents of the \texttt{other} peril so we avoid interpreting this result. The perils with the least number of significant embeddings are \texttt{fire} and \texttt{W\&H}. One notices that wind and hail are a meteorological phenomena, making sense that geographic embeddings, build on socio-demographic variables, are not all significant. One concludes that geographic embeddings generate statistically significant regression coefficients, so they are useful for ratemaking models. This conclusion holds when deconstructing the claim frequency into individual perils.

%Insignificant
%Fire 6
%Total 2
%Theft 3
%SBU 5
%Water 5
%W\&H 6 
%Other 3

\subsubsection{Predicting in a new territory}

\begin{figure}[ht]
	\centering
	\begin{subfigure}[b]{0.49\textwidth}
		\centering
		\includegraphics[width = \textwidth]{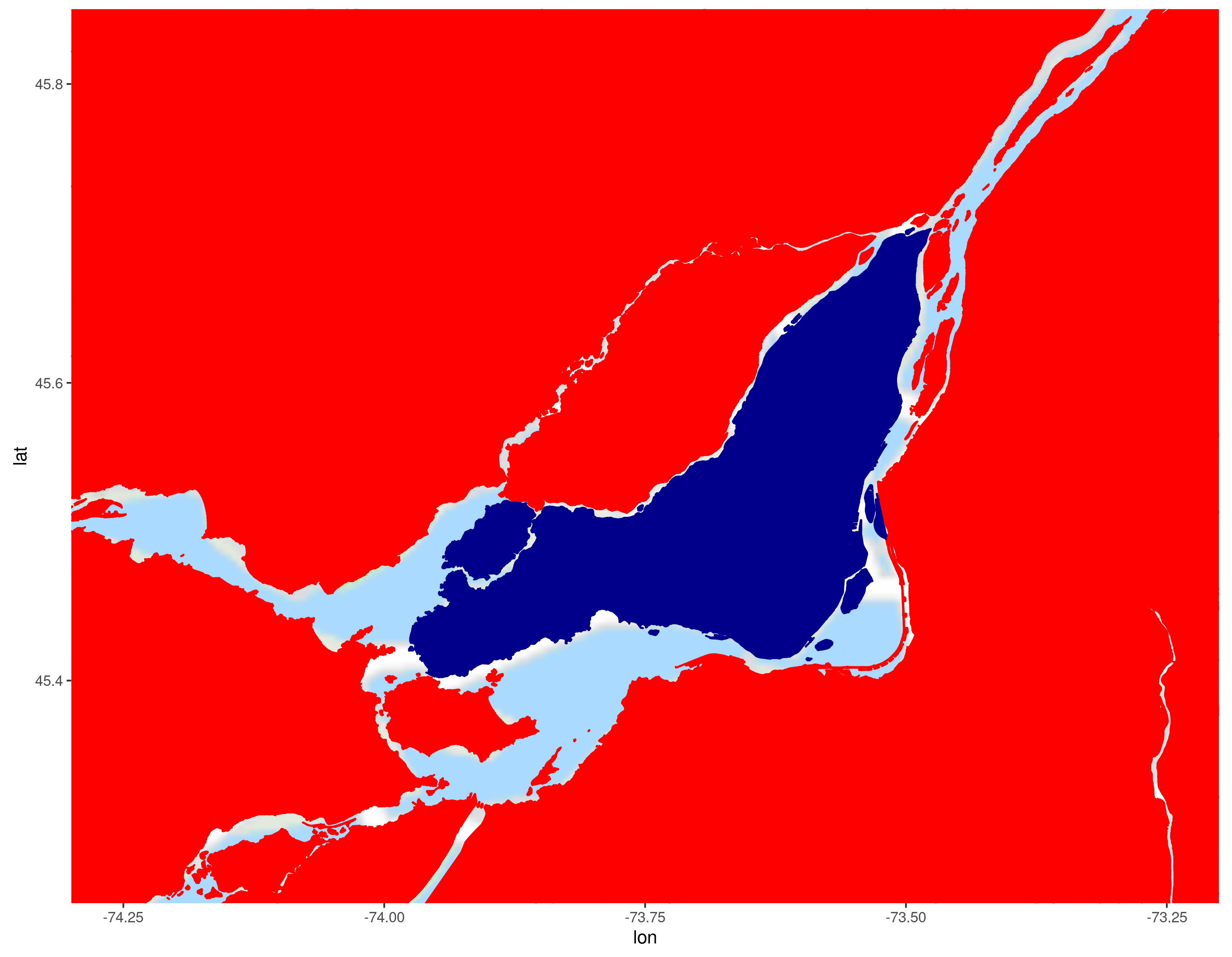}
		\caption{Zoomed in}
		\label{fig:map-small}
	\end{subfigure}
	\begin{subfigure}[b]{0.49\textwidth}
		\centering
		\includegraphics[width = \textwidth]{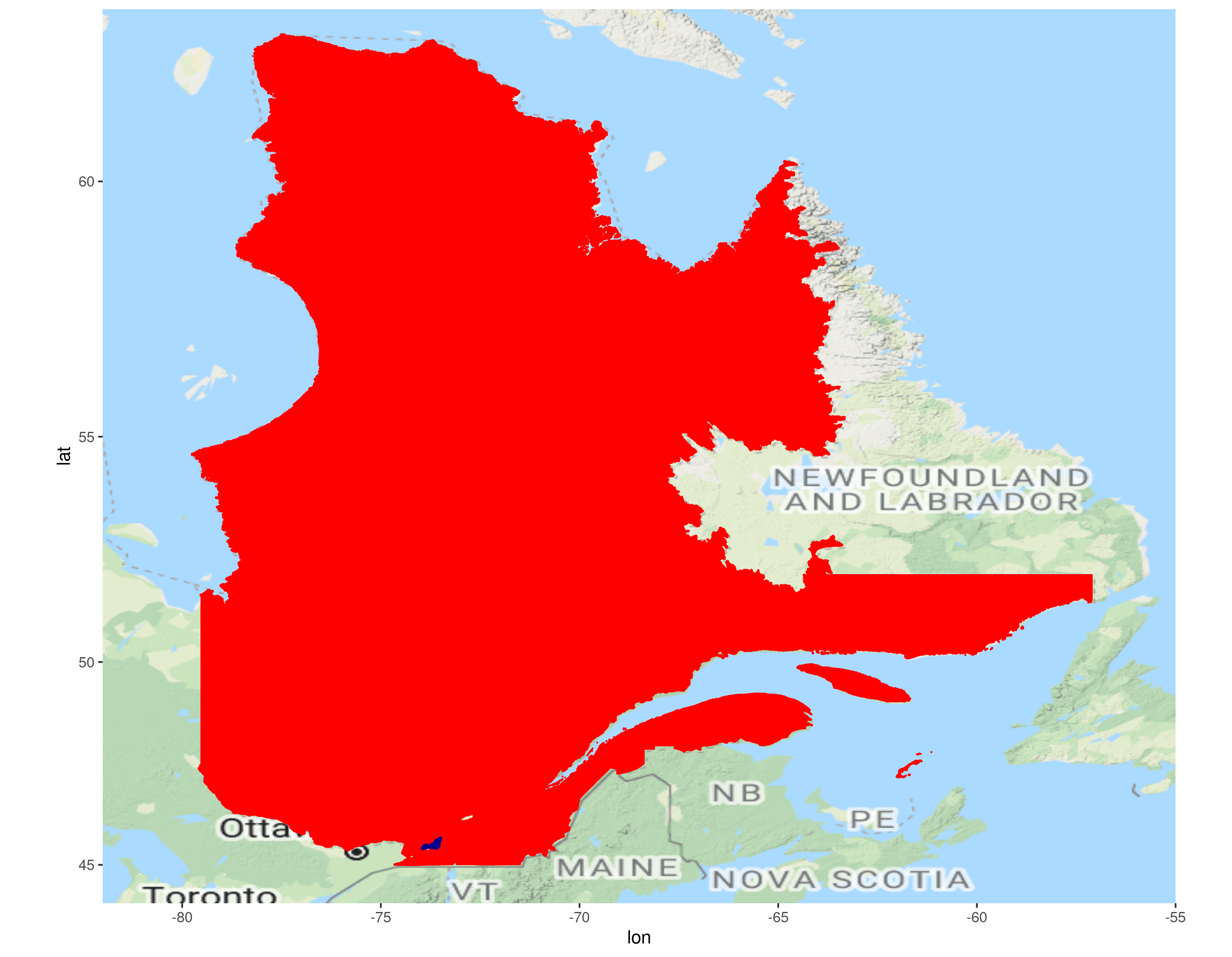}
		\caption{Zoomed out}
		\label{fig:map-large}
	\end{subfigure}
	\caption{Blue (darkest): Island of Montréal, red (lightest): Rest of Québec.}
	\label{fig:map-split}
\end{figure}

In the third comparison with the home insurance dataset, we determine if embeddings can predict geographic risk in a territory with no historical losses. On the one hand, we train a GLM with embeddings on a dataset with no historical losses from a sub-territory. On the other hand, we train a GLM with embeddings or a GAM with bivariate splines on a dataset with observations exclusively from that sub-territory. Then, we compare the performance of both approaches. To study this question, we split the dataset into two territories; Figure \ref{fig:map-split} illustrates an example for the island of Montréal.
The polygon in blue represents the island of Montreal (see Figure \ref{fig:map-small} for a focused look on the island), and in red is the remainder of the province (see Figure \ref{fig:map-large} for the entire province, with Montréal in blue at the bottom left). We train a Poisson GLM with embeddings for the two datasets: model OOT (out of territory) trains on all observations in the province except Montréal (red area), while model WT (within the territory) trains on observations in Montréal (blue area). Therefore, model OOT never saw any observations from Montréal. We also train a Poisson GAM with $k = 10$ as a baseline. All models use 2007 to 2010 calendar years for training data. Then, we compare the performance of the models on Montréal in 2011 (blue territory). The test deviance for model OOT is 14364, while the test deviance for model WT is 14383. In Table \ref{tab:villes}, we reproduce the results for the population centers in Québec with a population of over 100 000 and present the deviance computed on 2011 data.
%      2007     2008     2009     2010     2011
%A 12378.29 20473.24 16966.02 16670.30 14384.56
%B 12367.98 20415.77 16959.81 16702.40 14382.85
%R 12349.50 20697.43 17001.53 16669.55 14364.16

\begin{table}[ht]
	\centering
	\begin{tabular}{@{}llll|llll@{}}
		\hline
		Population center                             & WT            & OOT            & GAM   & Population center & WT   & OOT           & GAM  \\
		\cmidrule(r){1-4}\cmidrule(r){5-8}
		Montréal & 14383         & \textbf{14364} & 14400 & Sherbrooke        & 2783 & \textbf{2754} & 2782 \\
		Québec                                        & 3803          & \textbf{3787}  & 3801  & Saguenay          & 961  & \textbf{958}  & 963  \\
		Laval                                         & \textbf{4043} & 4046           & 4043  & Levis             & 2505 & \textbf{2500} & 2500 \\
		Gatineau                                      & 6495          & \textbf{6406}  & 6495  & Trois Rivières    & 2961 & \textbf{2952} & 2961 \\
		Longueuil                                     & \textbf{3184} & 3203           & 3185  & Terrebonne        & 2674 & \textbf{2656} & 2677 \\ \bottomrule
	\end{tabular}
	\caption{Test deviance with different training datasets by population centers}\label{tab:villes}
\end{table}

The models trained with OOT typically have smaller test deviance, so generalize better. For a few population centers, it is more beneficial to use data from the actual territory, but GLMs with geographic embeddings still outperform GAMs. These results lead to an important conclusion: to predict geographic risk in a sub-territory of a dataset, it is more advantageous to use geographic embeddings on a model trained with a larger quantity of losses, no matter where these losses occur, than to train a model using data from the sub-territory exclusively. When performing territorial ratemaking with traditional methods like GAMs, one typically focuses on one sub-territory at a time, thus ignores the remainder of the training data. On the other hand, OOT models use much larger datasets than the WT or GAM models, and this increased volume is more beneficial to improve the predictive performance than only studying the information from the territory of interest. This means that a model using geographic embeddings on a large quantity of information can improve geographic prediction in a territory with no historical losses better than if one had historical losses. In practice, one would not intentionally omit the observations in a sub-territory to predict the geographic risk: we construct this comparison to show that GLMs with embeddings can predict the geographic risk in territories with no historical losses. 

	% !TeX spellcheck = en_US
\section{Conclusion and Discussion}\label{sec:conclusion}

This paper shows that geographic models (bivariate splines) are unnecessary to capture the geographic distribution of risk. Instead, we can first capture the population's geographic distribution within a geographic representation learning model that outputs a vector of values. Using this vector in a simpler model such as a GLM predicts the geographic risk of home insurance losses more accurately than explicit geographic GAMs and enables predicting losses in territories with no historical losses. 

The ideas of Sections \ref{sec:crae} and \ref{sec:implementation} require much technical knowledge on neural networks and geographic models. These skills are outside of the typical actuary's knowledge, although this is rapidly changing. One main advantage of the geographic embedding approach is that another can use them within any predictive model after constructing the geographic embeddings. In both applications of Section \ref{sec:application}, we use the same set of embeddings for postal codes; we did not need to create specific representations for specific applications since we keep the geographic embeddings general. This could lead to a new relationship between actuaries, other domain experts and data scientists: one of a streamlined process of creating representations of novel sources of data, followed by actuarial modeling within existing models, akin to an assembly line. 

Training the embeddings in Section \ref{sec:implementation} required no knowledge of the insurance industry nor insurance data. The resulting embeddings are a geographic compression of socio-demographic information from census data. For this reason, the same set of embeddings could also be used for other tasks in domains where socio-demographic information could be useful, including life insurance, urban planning, election forecasting and crime rate prediction.

Note that embeddings do not entirely replace geographic models: they provide a simple way of compressing socio-demographic data into a vector. In some cases, GLMs using geographic embeddings are useful enough to avoid complex geographic models. In others, they provide a simple vector of features such that ratemaking models perform well. 

We offer two justifications of why we believe geographic embeddings work so well. First, the embeddings geographically capture geographic patterns that represent census information on a location and its neighbors. Since human-generated risks are closely related to human information, the geographic distribution of census embeddings will likely be related to the geographic distribution of risks. Therefore, a model using the census embeddings, even one as simple as a GLM, will be capable of modeling geographic risk. Second, insurance data is inherently noisy, so a model requires a high volume of historical losses to generate credible predictions. To build a geographic model, one can only use a subset of data to learn the geographic effect: the size of territories must be large enough to achieve credibility. When using geographic embeddings, the same features are used for the entire model, so the regression parameters associated with the geographic embeddings achieve credibility faster by using fewer parameters. One can interpret these two claims as follows: observations from Montréal can help predict the geographic risk in Québec since they both use the same geographic features. If an embedding dimension in Montréal generates higher risk, a GLM will capture this effect. An observation in Québec with a similar embedding dimension will predict a similarly higher risk, making sense because similar embedding dimensions imply similar census data, meaning the same type of individuals inhabits both places so that the socio-demographic risks will be similar. 

We proposed one intrinsic evaluation for geographic embeddings. For more, we could look at the field of natural language processing (NLP), where embeddings are used for many applications \cite{mikolov2017advances}. Researchers in this field have developed a set of tasks to evaluate the quality of embeddings. A common intrinsic evaluation task for natural language processing is computing a similarity distance between different two word embeddings and comparing the results with human similarity ratings from datasets like wordsim353 \cite{finkelstein2002placing}. One can also compare the similarity between different locations. For example, select three coordinates: two from similar territories and one from a dissimilar one. Embeddings intrinsically make sense if the similar territories have a smaller cosine distance than with the dissimilar embedding. This task depends on human interpretation, so without a survey asking many participants to rate the similarity of various territories, this intrinsic evaluation is not reliable for geographic embeddings. Future work on geographic embeddings could perform such a survey for an additional intrinsic evaluation. 

    \section{Acknowledgments}

The first author gratefully acknowledges support through fellowships from the Natural Sciences and Engineering Research Council of Canada (NSERC) and the Chaire en actuariat de l’Université Laval. This research was funded by NSERC (Cossette: 04273, Marceau: 05605). We thank Intact Financial Corporation along with the support and comments from Frederique Paquet, Étienne Girard-Groulx and Étienne Bellemare-Racine. We train geographic embeddings in PyTorch \cite{paszke2017automatic}. For statistical models, we use \cite{team2013r} and we fit GAMs with the mgcv packages \cite{wood2012mgcv}.

	\bibliographystyle{apalike}
	\bibliography{ref-clean}
    
    \appendix
    \newpage
    % !TeX spellcheck = en_US
\section{Canadian census details}\label{sec:census-detail}

In this section, we present the categories of variables that are available for every FSA in the Canadian census data. We exclude the variables from the categories denoted with an asterisk (*).

\begin{multicols}{2}
\begin{enumerate}
    \item Population and dwellings
    \item Age characteristics
    \item Household and dwelling characteristics
    \item Marital status
    \item Family characteristics
    \item Household type
    \item Knowledge of official languages*
    \item First official language spoken*
    \item Mother tongue* 
    \item Language spoken most often at home* 
    \item Other language spoken regularly at home* 
    \item Income of individuals in 2015
    \item Income of households in 2015
    \item Income of economic families in 2015
    \item Low income in 2015
    \item Knowledge of languages*
    \item Citizenship*
    \item Immigrant status and period of immigration*
    \item Age at immigration*
    \item Immigrants by selected place of birth*
    \item Recent immigrants by selected places of birth*
    \item Generation status*
    \item Admission category and applicant type*
    \item Aboriginal population*
    \item Visible minority population*
    \item Ethnic origin population*
    \item Household characteristics
    \item Highest certificate, diploma or degree
    \item Major field of study - Classification of Instructional Programs (CIP) 2016
    \item Location of study compared with province or territory of residence with countries outside Canada*
    \item Labour force status
    \item Work activity during the reference year
    \item Class of worker
    \item Occupation - National Occupational Classification (NOC) 2016
    \item Industry - North American Industry Classification System (NAICS) 2012
    \item Place of work status
    \item Commuting destination
    \item Main mode of commuting
    \item Commuting duration
    \item Time leaving for work
    \item Language used most often at work
    \item Other language used regularly at work
    \item Mobility status - Place of residence 1 year ago
    \item Mobility status - Place of residence 5 years ago
\end{enumerate}
\end{multicols}

\end{document}